\documentclass[12pt]{article}

\usepackage{amsmath,amssymb,graphicx} 
\usepackage{epsf}
\usepackage{pstricks}
\usepackage{cite}

\newcommand{\beq}{\begin{eqnarray}}
\newcommand{\eeq}{\end{eqnarray}}

\newcommand{\centeron}[2]{{\setbox0=\hbox{#1}\setbox1=\hbox{#2}\ifdim

\wd1>\wd0\kern.5\wd1\kern-.5\wd0\fi
\copy0

\kern-.5\wd0\kern-.5\wd1\copy1\ifdim\wd0>\wd1
                                       \kern.5\wd0\kern-.5\wd1\fi}}
\newcommand{\ltap}{\>\centeron{\raise.35ex\hbox{$<$}}
                               {\lower.65ex\hbox{$\sim$}}\>}
\newcommand{\gtap}{\>\centeron{\raise.35ex\hbox{$>$}}
                               {\lower.65ex\hbox{$\sim$}}\>}

\newcommand\ZZ{\hbox{\zfont Z\kern-.4emZ}}
\font\zfont = cmss10 

\newcommand{\met}{\mbox{${\rm \not\! E}_{\rm T}$}}
\def\gappeq{\mathrel{ \rlap{\raise.5ex\hbox{$>$}}
                      {\lower.5ex\hbox{$\sim$}}  } }
\def\lappeq{\mathrel{ \rlap{\raise.5ex\hbox{$<$}}
                      {\lower.5ex\hbox{$\sim$}}  } }
\begin{document}
\begin{titlepage}
\begin{flushright}
{\tt hep-ph/0601124}
\end{flushright}

\vskip.5cm
\begin{center}
{\LARGE Top Partners at the LHC: Spin and Mass Measurement \\
\vspace{.2cm}}

\vskip.1cm
\end{center}
\vskip0.2cm

\begin{center}
{\bf
Patrick Meade and Matthew Reece}
\end{center}
\vskip 8pt

\begin{center}
{\it Institute for High Energy Phenomenology\\
Newman Laboratory of Elementary Particle Physics\\
Cornell University, Ithaca, NY 14853, USA } \\
\vspace*{0.3cm}
{\tt  meade, mreece@lepp.cornell.edu}
\end{center}

\vglue 0.3truecm

\begin{abstract}
\vskip 3pt \noindent If one takes naturalness seriously and also
assumes a weakly coupled extension of the Standard Model (SM) then
there are predictions for phenomenology that can be inferred in a
model independent framework.  The first such prediction is that
there must be some colored particle with mass $\mathcal{O}$(TeV)
that cancels the top loop contribution to the quadratic divergence
of the Higgs mass.  In this paper we begin a model independent
analysis of the phenomenology of this ``top partner," $t'$.  We
make one additional assumption that it is odd under a parity which
is responsible for the stability of a WIMP dark matter candidate,
$N$. We focus on three questions to be explored at the LHC:
discovery opportunities, mass determination, and spin
determination of this top partner. We find that within a certain
region of masses for the $t'$ and $N$, $t'\bar{t'}$ is easily
discovered in the $t\bar{t}+2N$ decay with the tops decaying fully
hadronically.  We show that without having to rely on other
channels for new physics that for a a given $t'$ spin the masses
of $t'$ and $N$ can be measured using kinematic information (e.g.
average $\met$ or $H_T$) and total cross section. A degeneracy due
to the spin remains, but with several hundred fb$^{-1}$ of
luminosity we demonstrate potentially useful new methods for
determining the $t'$ spin over a wide range of masses. Our methods
when could be useful for distinguishing supersymmetric and
non-supersymmetric models.
\end{abstract}

\end{titlepage}

\newpage


\section{Introduction}
\label{sec:intro}
\setcounter{equation}{0}
\setcounter{footnote}{0}

In building models of physics beyond the Standard Model (SM),
naturalness has been a key motivating guideline. The Higgs boson
of the SM receives radiative corrections to its mass from loop
diagrams involving SM particles. These corrections are
quadratically divergent, implying that either new physics cancels
them (naturalness) {\em or} the bare mass of the Higgs is
finely-tuned. The canonical example of weakly coupled natural new
physics is supersymmetry, in which the quadratic divergences are
cancelled by superpartners of known particles having spins
differing by 1/2. However, in recent years there have been many
new ideas for implementing naturalness in a weakly coupled
framework (at least at the TeV scale), among them little Higgs models
\cite{LHoriginal}.

Experimental consequences of models implementing naturalness have
been explored to some extent, but the space of possible models is
large.  The space of models is also augmented by the fact that a
given model (for instance the MSSM) can also have a large set of
freely chosen parameters. Since the space of models and their
parameters is so large, many phenomenological studies can become
mired in the model dependent consequences of a particular
implementation of new physics. In this paper we advocate an
alternative direction: we wish to explore {\em model independent}
LHC phenomenology motivated by naturalness. This allows us to
focus in on key signatures that can be motivated from naturalness
without becoming stuck in the framework of a particular model.
Additionally we focus only on the LHC to investigate how much
physics information can be gleaned from its results alone,
independent of the abilities, or lack thereof, of any future
machine.

We begin by looking at the key ingredients in a natural extension
of the SM.  Assuming that an SM-like Higgs exists and physics
beyond the SM is weakly coupled at the TeV scale, the first
expectation from naturalness is an enlarged ``top sector." The top
loop in the SM is the largest contribution to the Higgs mass
quadratic divergence. Thus there must be some new particle(s)
constrained by symmetry to have couplings related to those of the
top, which cancel this loop. For instance in supersymmetry there
are scalar tops (``stops").  In Little Higgs theories there are
fermionic top partners. We call the generic top partner $t'$. One
could continue adding new particles based on naturalness, such as
partners of the gauge bosons and Higgs, but we will focus now
solely on the top sector.

We assume that the $t'$ that cancels the top loop is in the
fundamental representation of SU(3), as in all examples
we are aware of, so that
some symmetry can relate it to the $t$. Thus the $t'$ couples to
gluons and will be produced at the LHC in the processes
$gg\rightarrow t'\bar{t'}$ and $q\bar{q} \rightarrow t'\bar{t'}$.
Assuming no new particles or in scenarios where
the $t'$ decays to other particles beyond the SM but no stable new
particle exists, the resulting decay of the $t'$ will then solely
end in SM particles. Such SM decay modes will be relatively easy
to find at the LHC, as one can apply various cuts and build
invariant masses to find a resonance (similarly to how the top
itself is studied.)

On the other hand, if one has decays to SM particles plus a stable
neutral invisible particle (as in SUSY), the situation is much
more challenging.   With a stable invisible particle escaping
detection one cannot simply construct invariant masses. On the
other hand, requiring large $\met$ can dramatically cut back
Standard Model backgrounds. This scenario is also well-motivated
both from providing a dark matter candidate and from various
precision constraints. In typical models of physics beyond the SM,
four-fermion operators or electroweak oblique corrections are too
large without a parity forbidding the largest contributions.
Therefore beyond naturalness we will further assume, for now, that
there is some conserved $\mathbb{Z}_2$ parity. Examples include
R-parity~\cite{Rparity}(or Matter Parity~\cite{MParity}),
T-parity~\cite{TParity}, and KK-parity~\cite{UED}. The lightest
particle charged under this parity is stable. We will call the
stable particle the LPOP (``lightest parity-odd particle"), since
we do not assume a particular model for the parity.

For this paper we focus on a minimal scenario which is very
plausible for the LHC (if naturalness has any role in TeV scale
physics): there is some parity-odd heavy top $t'$, of undetermined
spin, decaying to the usual top quark $t$ and the LPOP $N$.
Because of the odd parity we must pair-produce the $t'$, so the
collider signature will be $t\bar{t} + \met$. Of course, a truly
model-independent approach would have undetermined couplings for
every decay of the $t'$ that does not violate a symmetry, so we
would also consider, for instance, $t'\rightarrow cN$ or  $t'
\rightarrow be^+\nu_{e}N$ (without the constraint $m_{be\nu} =
m_t$, e.g. through a $W'$). However, in most particular models the
$t' \rightarrow tN$ decay is dominant, or at least large. Since
the $t'$ and $t$ are closely related this is not surprising. Hence
we consider only the minimal scenario with decay $t' \rightarrow
tN$. As we will discuss, even when other decay modes are
available, the approach we outline can still be useful.

We should stress that we do {\em not} have particular models in
mind. We work from minimal effective Lagrangians that display a
signature that could be seen in any number of models (perhaps even
unnatural ones). Our study is {\em signature based}: we build very
general effective Lagrangians that generate $t\bar{t} + \met$,
under the assumption that it arises from $t'\bar{t'} \rightarrow
t\bar{t}+2N$, and focus on pinning down properties of the new
particles at the LHC.   Part of our motivation for such a study is
provided by the observation that certain models, like universal
extra dimensions~\cite{BosonicSUSY} or little Higgs models with
T-parity~\cite{jaypatrick,CLW}, can to some extent ``fake"
supersymmetric spectra unless one can measure spins. Some recent
papers have addressed how to distinguish spins/models in
particular cascade decays~\cite{IsItSUSY,Barr,SmillieWebber,NewBarr}. We think
that more studies along these lines are important for having a
realistic sense of how well the LHC can discriminate among models
and for building a preliminary toolkit of ideas and techniques for
analysis.  Another motivation is that it is quite plausible the
correct model of TeV scale physics has not yet been written down
but nevertheless naturalness could be a key ingredient in how
nature works. By parameterizing one of the most logical
possibilities for naturalness in terms of these minimal effective
lagrangians we can still analyze the possibilities and develop new
techniques for the LHC without requiring all the details of
nature's choice for the TeV scale.

In this paper we attempt to answer the following questions:
\begin{itemize}
\item Can the signature be observed at high significance over the SM backgrounds?
\item How well can we determine the masses of the $t'$ and $N$?
\item Can we devise an algorithm for measuring the spin of the $t'$ or the $N$?
\end{itemize}
As we will show, with several years of high-luminosity running all
of these except perhaps the spin determination for the $N$ should
be possible at the LHC. In the all-hadronic decay mode, a high
signal-to-background ratio can be achieved over a wide range of
masses.  Kinematic variables like average $\met$ or $H_T$ (see
Sec.\ref{sec:kin} for definitions) give some indication of the
mass splitting between the $t'$ and $N$, while the total cross
section determines the $t'$ mass for a given spin of $t'$. This
leaves a degeneracy. For the same cross section and kinematic
averages, the case of scalar $t'$ has a lower mass and so a
higher overall boost. We show that this allows pseudorapidity distributions
of the $t$ and $\bar{t}$ to be used, given enough luminosity, to
determine the $t'$ spin for a wide range of masses.

The rest of the paper is organized in the following way.  In
Section~\ref{sec:framework} we set up the details of the model
independent framework that we analyze as well as giving examples
of existing models included within our setup. In
Section~\ref{sec:signal} we examine the discovery possibilities
for the signal at the LHC.  In Section~\ref{sec:massdet} we
analyze the possibilities of mass determination and point out a
degeneracy due to spin that is often overlooked.  We then attack
the issue of spin determination in Section~\ref{sec:spin} where we
present new asymmetries and we use pseudorapidity
correlations to determine the spin of the $t'$.  We then discuss
in section~\ref{sec:nonSMbg} the impact of our model independent
study when considered in the context of existing models.  We
conclude by discussing future research directions for this
comparatively rare type of model independent analysis.

\section{Model Independent Framework}
\label{sec:framework} \setcounter{equation}{0}
\setcounter{footnote}{0}

In this section, we wish to set up the framework for what we
study. We are trying to address the question of what is a
reasonable first signature of new physics at the LHC in a model
independent manner, but we must construct some effective
Lagrangian to work with. We have established that our particle
content is a heavy top partner $t'$ and the LPOP $N$ with a decay
$t' \rightarrow tN$.  In principle we could fix the quantum
numbers of the $t'$ by investigating every possible mechanism of
cancelling the quadratic divergence of the top quark in the SM.
While this would allow for a completely model independent study
there presumably could always be a loophole for the cancellation
that we have missed.  There are many existing models with top
partners that cancel (at least to one loop)
the quadratically divergent contribution of
the top quark to the Higgs mass, for instance SUSY or Little
Higgs. What we find from this is that there are two reasonable
possibilities for the top partner spin: it could be either a spin
$1/2$ fermion or a scalar~\footnote{More exotic possibilities
could perhaps occur if there were some composite resonance but we
will avoid these cases since there isn't an obvious symmetry to
relate the couplings. In this paper for the spins of the particles
we will restrict ourselves to the case of spin 1 and less.}. While
this may be viewed as introducing some model dependence we will
{\em not} restrict ourselves to the parameters in these models.
The spins implied by the various existing mechanisms for
naturalness are only used as a possible starting point.

Once the spin of the $t'$ is fixed, the vertex for the decay
$t' \rightarrow tN$ fixes the possible spins of
the $N$. In the case that the $t'$ is a scalar $N$ must
be a fermion. On the other hand if the $t'$ is a fermion it
implies that $N$ could either be a scalar, spin $0$, or vector,
spin $1$, particle.

At this point we need a way to fix the SM gauge quantum numbers of
the $t'$ and $N$.  As was mentioned before we have already
demanded that the $t'$ is in the fundamental representation of
$SU(3)$. However, the electroweak quantum numbers have
not been fixed. Since the LPOP is a neutral stable
particle the vertex $t'tN$ also fixes the electric charge of the
$t'$ to be the same as the top quark.  Hence we only must
determine the representation of $SU(2)$ that the $t'$ candidate
resides in. This could in principle be fixed by examining the
various mechanisms for cancelling the quadratic divergences of the
Higgs mass and what the vertex with the Higgs implies for the
$SU(2)$ quantum numbers. Generically in existing models there are
cases where there are top partners which can be both singlets and
doublets of $SU(2)$. Since we are trying to present a new
framework for studying physics beyond the SM, to limit the scope
of our study we will only analyze the case where both the $t'$ and
$N$ are singlets of $SU(2)$. This does not mean we assume there
are no doublets, merely that they are heavy enough and with
small enough mixing that they do not influence the results.
This choice avoids the issues of
introducing a partner of the bottom quark or additional particles
associated with the $N$ (for instance charginos or other gauge
bosons). In Section~\ref{sec:nonSMbg} we will further discuss the
issues relevant to our study created from introducing new
particles.

We now summarize the particle content for our study: a singlet
$t'$ which is either a scalar or fermion with mass $m_{t'}$, and
the LPOP singlet $N$, with mass $m_N$, which is a fermion in the
case of a scalar $t'$ and a scalar or vector in the case of a
fermionic $t'$. Because of our assumptions the coupling to the SM
top quark takes the form
\begin{equation}\label{rhtcoup}
\mathcal{L}_{tt'N}\sim g_{t'N} (t' \bar{t}_R N + h.c.).
\end{equation}
In principle we could relax the assumption of having only a light
singlet $t'$ and it could mix with a heavier doublet partner of
the top leading to a coupling of the form
\begin{equation}\label{mixcoup}
\mathcal{L}_{tt'N}\sim t'\bar{t} (g_{t'N_L} P_L + g_{t'N_R} P_R)N
+ h.c.,
\end{equation}
where $P_{L,R} = (1 \pm \gamma_5)/2$ and we have omitted the
details of the Lorentz structure that depend on the spins of $t'$
and $N$.  This resembles more closely the situation in SUSY where
$\tilde{t}_L$ and $\tilde{t}_R$ mix to form $\tilde{t}_1$ and
$\tilde{t}_2$.  So as not to complicate our scenario any further
we will assume that the coupling is of the form (\ref{rhtcoup}). We
will discuss the effects of this assumption in Sections \ref{sec:massdet}
and \ref{sec:spin}.
Thus for our study there will naively be three
parameters separate from the choice of spin, $m_{t'}$, $m_N$, and
$g_{t'N}$. Since the $t'$ only has one decay channel available to
it, calculating its decay width properly means that $g_{t'N}$ is
not actually an additional parameter in our study (except for
small interference effects, important only when $m_{t'} \approx
m_t + m_N$).

\subsection{Model Dependent Realizations}

So far we have presented a model independent realization of what
one would expect from naturalness supplemented by a parity with a
neutral LPOP.  This realization is only a subset of possibilities
for a top partner but nevertheless a sensible starting point.  It
is important to note that this model independent framework that
we have laid out can also realize many existing models of physics
beyond the SM.  In most of the cases that we will discuss there
are additional other particles beyond our minimal set and their
impact will be discussed in Section~\ref{sec:nonSMbg}.

The case of a scalar $t'$ is obviously analogous to a
stop in the MSSM.  For our particular choice of quantum numbers
the scenario in the MSSM would correspond to $t'$ being a light
$\tilde{t}_1$ that was predominantly $\tilde{t}_R$, while the $N$
would correspond to the lightest neutralino, $\chi^0_1$, being
mostly Bino and the LSP (LPOP). A particular realization
where these are the lightest new particles in the MSSM is not the
most common region of parameter space.  However, it is a relatively
typical region of parameter space in mSUGRA to have the
$\tilde{t}_R$ being the lightest colored particle and a Bino-like
LSP.  While there may be other particles in the same mass range
relevant to the LHC, the decay $\tilde{t}_1\rightarrow t\chi^0_1$
typically has a sizeable branching fraction.

Having a fermionic $t'$ cancel the quadratic divergence to the
Higgs mass from the $t$ occurs within the framework of Little
Higgs models~\cite{LHoriginal}.  However since we are only
interested in the case of a parity odd $t'$ the relevant models
are those with a T-parity\cite{TParity}.
In the original models of Little Higgs with
T-parity the cancellation of the top loop divergence in the SM was
due to a parity even top quark partner but in addition there
was always a T-odd partner $t'_{-}$ which is
lighter\cite{jaypatrick}.  The decay of the $t'_{-}$ in the
Littlest Higgs with T-parity has only one channel
$t'_{-}\rightarrow tA_H$ where $A_H$ is the LTP (LPOP), a heavy
partner of the hypercharge gauge boson of the SM (up to small
$v/f$-suppressed mixings).  This is the
case of a $t'$ fermion and $N$ vector that we have laid out.  In
addition there has been a recent paper~\cite{CLW} where the
$t'_{-}$ was responsible for cancelling the divergence to the
Higgs mass, which is also directly realized in our framework.

The examples of specific models contained within our model
independent realization given so far, not surprisingly, are guided
by the assumption of naturalness.  However within our naturalness
motivated model independent framework it is also interesting to
note that we can accommodate models that do not necessarily
address the question of naturalness.  An example of this type is
UED\cite{UED} models, which have received considerable attention
as an example of a model which can be confused with
SUSY\cite{BosonicSUSY}.   In these models the SM propagates into
an extra dimension(s) and thereby KK partners of all SM particles
exist. These models have a discrete symmetry, KK-parity, that
provides an LKP (LPOP) that is typically the KK partner of the
photon. Therefore in these types of models the KK partner of the
top quark $t_{1R}$ decays to $t_R$ and the LKP $A_{1}$, and so is
also realized within our framework.

From the above examples we see that the model independent
realization we have chosen can obviously be applied to a host of
existing models.  Of course the caveat exists that in many models
there will be other non-SM backgrounds that would need to be
taken into account which will be discussed in Section~\ref{sec:nonSMbg}.
Notwithstanding the particular examples of our framework
containing existing models, we reiterate that this framework is
not simply model inclusive but model independent. The parameters
of our study $m_{t'}$ and $m_N$ are usually fixed in all examples
that we have given so far based on other parameters in the given
model.  In our study $m_{t'}$ and $m_N$ will be varied freely
within a certain range that {\em can not necessarily} be realized
in the above examples.  We thus look for the signature
$t\bar{t}+\met$ in regions that would not
necessarily be accessible by the
models existing so far, hence the distinction model independent.

\section{Discovery Opportunities}
\label{sec:signal} \setcounter{equation}{0}
\setcounter{footnote}{0}

In this section we will examine the discovery opportunities for
the signal $t\bar{t} + 2 N$ where the $N$ is neutral LPOP that
escapes the detector.  Our signal has several possible channels
that we could look at depending upon how the $W$ gauge bosons from
the top end up decaying.  Comparing the leptonic or semileptonic
channels versus the hadronic channel, we see that in a leptonic
channel there is already a source of $\met$ (from neutrino(s)) in
the SM while in the hadronic channel there is no SM contribution
to the $\met$. Thus the largest SM irreducible background in the
leptonic case would be $t\bar{t}$ while for the hadronic channel
it would be $t\bar{t}Z$ where $Z\rightarrow
\nu\bar{\nu}$\footnote{There are many additional sources of
backgrounds that one must consider, such as W+jets or Z+jets,
different decays of a $t\bar{t}$ pair with additional jets
combined with incorrect particle IDs, energy mismeasurement, and so
on.  It turns out after applying our cuts and taking into account
various efficiency considerations the backgrounds just listed are
smaller. We will give our estimates of the various backgrounds in
Section~\ref{sec:backgrounds}.}. Therefore the backgrounds are
much smaller in the hadronic channels than in the semileptonic
channel. It is also important to note that not only are the
backgrounds smaller in the hadronic channel, there is more
kinematic information as well.  In the leptonic channels the
additional source of $\met$ from the neutrinos reduces the amount
of kinematic information about the original $t\bar{t}$ pair, whereas in
the hadronic channel the momenta of the top quarks can be fully
reconstructed. For these reasons we will focus on the hadronic
channel for our signal.

To study this signal we implement the effective framework
discussed in Section~\ref{sec:framework} using the MadGraph
software~\cite{Madgraph}.  The coupling $g_{t'N}$ in
(\ref{rhtcoup}) was set to the electric charge $e$.  However, as
discussed in Section~\ref{sec:framework}, this coupling is not
really a free parameter since we have assumed there is only one
decay channel.  Since we have assumed that the $t'$ is in the
fundamental representation of $SU(3)$ we implement a single gluon
coupling for a fermionic $t'$ and both a single and double gluon
vertex for the scalar $t'$.  In calculating all processes that
contribute to $t\bar{t} + 2 N$ we use the parton distribution
function CTEQ6L1~\cite{PDF}.  The renormalization and
factorization scales used for our signal will be the mass of $t'$.
The factorization and renormalization scale dependence of our
results could be minimized by calculating the effects beyond
leading order, but for now we refer to the NLO QCD corrections to
$t\bar{t}$ production where $m_t$ is a reasonable choice of
scale~\cite{Beneke:2000hk}. It is important to note that MadGraph
is only a tree level matrix element generator. For our study we
will compare tree level signal and background cross sections.  The
QCD corrections will tend to increase both our signal and
background as well as pin down the factorization and
renormalization scales.  This increase should be an order one
effect and could be absorbed into a K-factor but will not affect
our results in any qualitative manner in the bulk of our parameter
space. Future studies taking into account more detailed issues
could investigate this further but it is a reasonable starting
point to compare tree level signal and background to come up with
new techniques.

We will split the discussion of discovery opportunities into
several parts. We will begin by giving the definitions of all
kinematic variables used in our study.  We then describe the cuts
used in analyzing both signal and background.  Before discussing
the various backgrounds we will collect the efficiencies relevant
to the process studied.  We then give the background cross
sections taking into account our cuts and the efficiencies.  To
demonstrate the discovery opportunities we then look at both the
usual measure of signal to root background, and then the signal to
background ratio.

\subsection{Definitions of Kinematic Variables}\label{sec:kin}

To discriminate signal from background, we wish to apply a set
of kinematic cuts on certain variables. These variables also will prove useful in
determining the masses of the $t'$ and $N$, as we will discuss in
Section \ref{sec:massdet}.
Here we collect the definitions of the kinematic
variables we will be considering:

\begin{itemize}
\item $|\met|$: At the parton level we define $|\met|$ as the
length of the transverse momentum vector constructed from the
vector sum of transverse momenta of all neutrinos and the
$N$'s.\footnote{In practice, one cannot define $\met$ in this way.
Instead $\met$ is defined to be opposite to the vector constructed
from summing $E_T$'s of all visible objects. More precisely, $E_T$
of an object is normalized by the energy deposited in the
calorimeter, and the direction is obtained from the location of
the energy deposit relative to the event vertex. For muons one
would use $p_T$, as the momentum is well-measured from the track
but the muon does not deposit all of its energy in the
calorimeter.} \item $H_T$: $H_T$ is defined as the {\it scalar}
sum of transverse momenta of the reconstructed objects. More
precisely, $H_T$ is $|\met|$ plus the sum of $|E_T|$ for all jets,
electrons, and taus, plus the sum $|p_T|$ of all muons. \item
$M_{eff}$: $M_{eff}$ has developed a reputation as a good measure
of the mass scale of the strongly interacting particle
\cite{ATDR,Meff}. It is defined to be the sum of $|E_T|$ for the
four highest-$|E_T|$ jets, plus the $|\met|$, where the
distribution is only used for events satisfying certain cuts: we
require $|\met| > \max(100~{\rm GeV}, 0.2 M_{eff})$, the highest
jet $|E_T| > 100$ GeV, at least four jets with $|E_T| > 50$ GeV,
and no muon or isolated electron with $p_T > 20$ GeV and $|\eta| <
2.5$. \item $M_{T2}(m_{N;ref})$: Proposed by Lester and
Summers\cite{MT2}, $M_{T2}(m_N)$ is a variable that would give us
$m_{t'}$ if the $N$ mass is known precisely. This works in much
the same way that, for instance, the $W$ mass can be obtained from
a histogram of the transverse mass constructed from an electron
and missing energy. It is a generalization of transverse mass
defined as
\begin{equation}
M^2_{T2} = \min_{{\bf p}^{N_1}_T + {\bf p}^{N_2}_T = \not{\bf
p}_{T}} \left[ \max\left[m_T^2({\bf p}^{t_1}_{T},{\bf
p}^{N_1}_{T}),m^2_T({\bf p}^{t_2}_{T},{\bf
p}^{N_2}_{T})\right]\right],
\end{equation}
where ${\bf p}^{t_1}_T$ and ${\bf p}^{t_2}_T$ are the transverse
momenta of the reconstructed top quarks (or more generally, the
relevant reconstructed particles), and $m_T$ is the transverse
mass of two particles defined to be valid for arbitrary masses.
Then $M_{T2}$ by construction never exceeds the mass of the heavy
top, provided $m_N$ is known. The minimization is over all ways of
splitting the missing transverse momentum among the two
disappearing particles. The definition can be recast in a form
that requires only a one-dimensional minimization, which is easier
to compute. For this and other details see the original
papers~\cite{MT2}. Since $m_N$ is unknown as well in our case, we
must be careful about the use of this variable. We supply an
``input" $N$ mass, $m_{N;ref}$, which might be quite different
from the actual $N$ mass. The histogram of $M_{T2}(m_{N;ref})$ has
a lower edge determined by $m_{N;ref}$ and a fairly sharp upper
edge. We denote this upper edge $M_{T2}^{max}(m_{N;ref})$.
Changing the input $m_{N;ref}$ will shift the position of the edge
but does not alter its existence.
\end{itemize}

\subsection{Cuts}\label{sec:cuts}

Since our signal is $t\bar{t}+\met$, we clearly need to
require some large amount of $\met$ to discriminate our
signal from the Standard Model $t\bar{t}$ background.
In this way we should be able to ensure the
dominant background is $t\bar{t}Z$.
Also, we have to ensure that the events we consider
will pass the relevant triggers. Requiring $\met > 100$ GeV
and one jet with $E_T > 100$ GeV should be sufficient
for this requirement \cite{ATLAStrigger}. We place a
conservatively hard cut of 40 GeV on the $E_T$ of all
relevant jets, to help reduce QCD backgrounds and to
guard against multiple interactions and initial- and final-
state radiation. We demand two tagged $b$ jets to ensure
that our SM backgrounds are mostly top-related, as opposed
to the much more common $W+jets$ and $Z+jets$ events.
Furthermore, to be sure that we are looking at events involving
$t\bar{t}$ we apply some mass reconstruction cuts.

To summarize, we use the following set of cuts:
\begin{itemize}
\item Two $b$-tagged jets and four other jets having $E_T > 40$ GeV.
\item At least one jet with $E_T > 100$ GeV.
\item $\met > 100$ GeV.
\item $|\eta| < 2.5$ for all jets.
\item $\Delta R > 0.4$ between any pair of jets.
\item The four non-$b$ jets split into two pairs reconstructing
to a $W$: $60~{\rm GeV} < M_{jj} < 100~{\rm GeV}$.
\item The two $W$s pair up with the two $b$ jets to reconstruct
to a top: $150~{\rm GeV} < M_{jjb} < 190~{\rm GeV}$.
\item $H_T > 500~{\rm GeV}$, where
$H_T = \met + \sum_{jets} |{\bf p}_T|$ (see below).
\end{itemize}

Additional cuts can be made to attempt to make a more pure sample.
Additionally a more sophisticated analysis for the reconstruction
of the top quark mass window candidates has been looked at in the
past~\cite{ATLASstop}. We believe our cuts are conservative enough
though that our background estimates will be reliable. Once the
mass scale for the signal is determined from methods presented in
Section~\ref{sec:massdet} the cuts can be tuned further to enhance
the signal.  Recently an analysis of the ability to measure the
$t\bar{t}Z$ couplings at the LHC was made in the
$Z\rightarrow\nu\nu$ all hadronic channel using similarly
conservative cuts~\cite{ttz}.  In this study it was claimed
$t\bar{t}Z$ could be seen above the SM backgrounds thus with the
typically larger $\met$ of our signal it is even more likely our
estimates are conservative enough or can be made even better with
a larger $\met$ cut.

\subsection{Efficiency Considerations}

Here we will consider a few issues that are very important for us
to understand how efficient we can be at identifying candidate
signal events. First, how often will we tag a true $b$-jet, and
how often will a non-$b$ jet be tagged? It is clear that QCD
multijet backgrounds can be large if we do not insist on having
$b$ jets. Second, how often will a hadronically-decaying $\tau$
lepton fake a jet, and how often will a true jet be reconstructed
as a hadronically-decaying $\tau$? Backgrounds with
$W\rightarrow\tau\nu$ can be troublesome, so we need some ability
to veto them. There are, of course, other efficiencies to
consider. Problems can arise from mismeasurement of energies, for
instance. We will leave a detailed consideration of those
difficulties to experimentalists; such issues will become much
more clear when the LHC is operating. However we can make some
preliminary estimates of how troublesome these effects could be.

In the case of $b$ jets, the efficiency for tagging a true $b$ jet
in the $p_T$ range we are looking at is claimed to be about $60\%$
with a factor of 100 rejection for light-quark or gluon jets and a
factor of 10 rejection for charm jets\cite{ATDR}. Thus we expect
approximately a 30\% efficiency for correctly $b$-tagging the
signal. (At high luminosity running the efficiency of $b$-tagging
is expected to decline to 50\% at the same rejection.)

For $\tau$ leptons,  the situation is troubling, as we have many
light quark jets that can potentially reconstruct as a $\tau$ and
cause us to veto good signal events. The $\tau$ decays
hadronically about 64\% of the time\cite{PDG}. At high $p_T$ a
hadronically-decaying tau will fake a jet 5\% of the time while
10\% of light-quark jets will reconstruct as a
$\tau$\cite{ATLAStau}. Imposing a $\tau$ veto on all our jets will
thus impose about a 60\% efficiency factor. The cited study does
not include separate rates for light-quark jets, charm jets, and
$b$ jets faking taus, but uses some large sample of $t\bar{t}$,
$b\bar{b}$, and $W$+jets. It does note that $b$ jets are less
likely to fake taus than other jets. We will assume that the
distribution of various types of jets in the samples studied is
similar to that in our signal. We will also assume that we need
not reject any $b$-tagged jets that fail the tau veto. We explain
below that rejecting {\em every} event with a $\tau$ candidate may
not be necessary, as we can use transverse mass information to
help distinguish signal from background. Furthermore, one would
like to have a $\tau$ ID likelihood that allows us a finer
resolution on the extent to which a jet is tau-like. Highly
tau-like jets could be cause for immediate veto, while moderately
tau-like jets could be kept or not depending on other selection
criteria.

To summarize, we will define $\epsilon_{b\tau}$ to be the
efficiency, in a given sample, of tagging all $b$ jets and passing
a tau veto. We will give an estimate of this quantity for each
sample, but such estimates are very preliminary. It is important
to keep in mind that fake rates and efficiencies are dependent on
flavor and on signal. When the LHC begins to operate one can gain
a better understanding of the efficiencies and fake rates involved
(studying the large signal of $t\bar{t}$ in the SM will probably
be useful). With all of these caveats, we estimate that for our signal,
the efficiency is $\epsilon_{b\tau} \approx 18\%$.

Another consideration is mismeasurement, which can be problematic
in two main ways. The first is that signal events can fall outside
the mass window cuts due to poor measurement of jet energies. A
good understanding of detector resolution is needed to study this.
One might attempt to rescale energies in suitable candidate events
so that the $jj$ masses near the $W$ mass are constrained to equal
the $W$ mass. A more detailed study of such possibilities is
necessary. In the worst case, we lose efficiency by some factor
$\epsilon_{recon}$. ATLAS simulations of $t\bar{t}$ show a mass
peak with a width of 13.4 GeV \cite{ATDR}, so our 15 GeV window on
each top could imply that we miss about half the events. However,
since we require {\em both} tops to decay hadronically and lie in
the mass window, the combinatorial background is likely to be
smaller than in the semileptonic case. Further, one might impose a
weaker window on one of the tops than on the other.

The second mismeasurement concern is that $\met$ can be
overestimated in hadronically-decaying $t\bar{t}$ events. The
$t\bar{t}$ cross section is so large that even if this happens
rarely, it could still lead to backgrounds as large as the others
we are considering. There are really two distinct issues here: the
first is that jet energies can be mismeasured, creating $\met$
where there is none. This should not be very likely; ATLAS studies
have indicated that missing $E_{T}$ resolution is well-described
by $\sigma_{\met} = 0.46 \sqrt{\sum E_T}$ at low luminosity and is
twice as bad at high luminosity~\cite{ATDR}. Thus measuring 100
GeV of $\met$ where there is none should happen rarely. The other
source of $\met$ overestimation is that semileptonically decaying
$b$ jets can contain a high $E_T$ neutrino. We have taken a first
look at this, using the PGS detector simulation \cite{PGS} on a
sample of $t\bar{t}$ events generated with Pythia \cite{Pythia}.
It appears that roughly 0.2\% of all fully-hadronic $t\bar{t}$
events have $\met > 100$ GeV. For comparison, note that
$t\bar{t}Z$ has a cross section of order 0.1\% that of $t\bar{t}$.
Thus we can't immediately discount the $t\bar{t}$ background with
$\met$ from semileptonic $b$ decays. However, if a large fraction
of the $b$ energy goes to a neutrino, it is unlikely that the
reconstructed $jjb$ mass will lie near the top mass, so we expect
the mass window cut to have much lower efficiency on the
$t\bar{t}$ sample than on the $t\bar{t}Z$ sample. To get a
definitive answer to this question, a study of all the backgrounds
with detector simulation is necessary. We expect that after all
cuts are applied it is not problematic.\footnote{We are aware of a
study in progress of all-hadronic physics at the LHC
(specifically, of the stau coannihilation region) with similar
$\met$ and $E_T$ requirements, which also reaches the conclusion
that SM $t\bar{t}$ background is low~\cite{StauLHC}.} At worst it
leads to an ${\cal O}(1)$ increase in background, which is only
detrimental at the high end of the mass range we consider. If this
problem proves to be much more severe than we estimate, one should
still be able to cut the extra background by simply imposing a
harder $\met$ cut of 150 GeV or 200 GeV. So long as $m_{t'}$ is
not too close to $m_N + m_t$, this will still keep a substantial
fraction of our signal events. The major drawback to using such
harder cuts is that one needs more integrated luminosity to build
up a clean signal sample for precision studies to determine spin.

One might also worry that our cut $\Delta R > 0.4$ between partons
could be insufficient to prevent the reconstructed jets from
overlapping, and that the resulting difficulty in reconstruction
could cause problems. We also ran our cuts with $\Delta R > 0.7$
and find that about 30\% of the $t\bar{t}Z$ background is kept. On
the signal, the efficiency of this cut is highly dependent on
$m_{t'}$ and $m_{N}$, but typically in the 20\% to 30\% range. At
high $m_{t'}$ mass the $S/B$ ratio will drop with the stricter
separation requirement. For our purposes we will keep the $\Delta
R > 0.4$ separation cut and assume that the jet algorithms are
capable of separating overlapping jets intelligently.

With all of these various efficiency issues we have outlined above
they must be revisited with a full detector simulation.  We have
attempted to be conservative in our estimates to give the reader
confidence in the fact that a full detector simulation of our
methods would be worthwhile.

\subsection{\label{sec:backgrounds} Backgrounds in the All-Hadronic Channel}

As we have noted, the largest background is the $t\bar{t}Z$
channel. We simulate this background with MadGraph. The quark
content of this sample is the same as in the signal, so we
estimate $\epsilon_{b\tau} \approx 18\%$. With the cuts listed
above, we obtain\footnote{We will use ``cross section" (denoted
$\sigma$) loosely to mean cross section times branching ratio
times relevant acceptances and efficiencies. That is, it is the
quantity that when multiplied by luminosity gives the expected
number of events passing all cuts. It includes our estimate of
$\epsilon_{b\tau}$ for a given sample.}:
\begin{equation}
\sigma(t\bar{t}Z) = 0.32~{\rm fb}.
\end{equation}

The next significant background to consider is $t\bar{t}j
\rightarrow \tau j j j b\bar{b}+\met$. Here the hadronic tau
decay, together with the extra jet, can sometimes mimic the two
jets from a $W$ decay. We simulate this background with MadGraph.
In this case we estimate $\epsilon_{b\tau} \approx 1.1\%$.
MadGraph only includes two hadronic decay modes of the $\tau$,
namely $\tau^- \rightarrow \pi^- \nu_{\tau}$ and $\tau^-
\rightarrow \rho^- \nu_{\tau}$. The $\rho$ events pass our cuts
slightly more often than the $\pi$ events, but the efficiency is
approximately the same. We will estimate that the other hadronic
channels pass our cuts as often as the $\rho$ channel, though of
course a more detailed analysis should simulate more decay modes
(including the three-prong modes). We estimate:
\begin{equation}
\sigma(t\bar{t}j\rightarrow \tau jjj b\bar{b}+\met) = 0.09~{\rm fb}
\end{equation}

In this and other $\tau$-related backgrounds where all the $\met$
is from the decay $W \rightarrow\tau \nu \rightarrow j+\met$, we
have the advantage that the transverse mass $m_T({\bf
p^{j}_T},{\bf \met}) < M_W$. Thus we do not have to reject every
candidate event in which one of the jets reconstructs as a
hadronic tau. We need only reject these events if they satisfy
this transverse mass constraint. Thus, we have a {\it pessimistic}
estimate of SM backgrounds (aside from possible
mismeasurement backgrounds as discussed above).

The other backgrounds rely on higher fake rates or accidental mass
reconstruction. They are small compared to $t\bar{t}Z$. We tabulate
them in Table \ref{tab:background}. Note that we have used Alpgen
\cite{Alpgen} to set a rough upper bound for the multi-jet processes.

\begin{table}[h]
\centering
\begin{tabular}{llll}
Channel  & Generator & $\epsilon_{b\tau}$ & $\sigma$ \\
 \hline
$ t\bar{t}Z~(Z\rightarrow\nu\nu)$ & MadGraph & 0.18 & 0.32 fb \\
 $t\bar{t}j \rightarrow \tau jjjb\bar{b}+\met$ & MadGraph & 0.011 & 0.09 fb \\
 $t\bar{t}jj \rightarrow \tau \tau j j b\bar{b}+\met$ & MadGraph & 0.0006 & $\lappeq 10^{-5}$ fb \\
 $Wb\bar{b}+3j~(W\rightarrow\tau\nu)$ & Alpgen & 0.01& $\lappeq0.009$ fb \\
 $W+5j~(W\rightarrow\tau\nu)$ & Alpgen & $3.5 \times 10^{-6}$  & $< 10^{-5}$ fb \\
 $Zb\bar{b}+4j~(Z\rightarrow\nu\nu)$ & Alpgen & 0.18 & $\lappeq0.022$ fb \\
  $Z+6j~(Z\rightarrow\nu\nu)$ & Alpgen & $6. \times 10^{-5}$ & $< 0.013$ fb \\
\end{tabular}
\caption{
Cross section (times branching ratios times efficiencies) for backgrounds to the
hadronic $t\bar{t}+2N$ decays, after all cuts have been applied. In the Alpgen
processes we have applied a subset of the cuts to get an upper bound. For
the $W+jets$ processes we have not decayed the $\tau$,
so our estimate is very rough (the$\met$ comes solely from the
$\nu_{\tau}$, but the hadronic tau decay involves other neutrinos).}
\label{tab:background}
\end{table}

\subsection{Results for Significance and Signal to Background}

Taking into account the various backgrounds and efficiencies we
will now present our results for the discovery opportunities at
the LHC.  We will quantify the discovery opportunity in the usual
way, after applying the cuts described in Section~\ref{sec:cuts}, of
\begin{equation}\label{eq:disc}
\mathrm{significance}=\frac{\mathrm{signal}}{\sqrt{\mathrm{background}}},
\end{equation}
where background is implicitly the number of background events
from summing all backgrounds weighted with the appropriate
efficiencies as found in Table~\ref{tab:background}, while signal is
the number of signal events multiplied by the efficiency defined
in Section~\ref{sec:signal}.  We will then discuss the signal to
background ratio, as it will be important for future
discussion as well as for giving another measure of discovery.

\begin{figure}[h]
\begin{center}
\includegraphics[width=8cm]{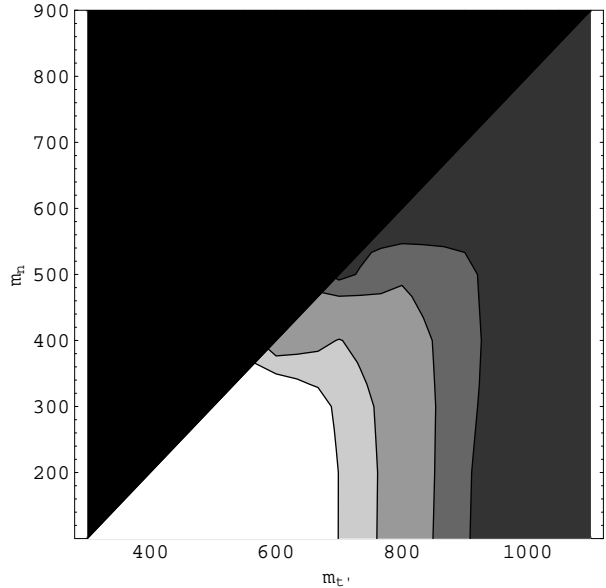}
\end{center}
\caption{Significance of signal for $t'$ fermion $N$ scalar for
$10\;\mathrm{fb}^{-1}$. The contours (from left to right)
represent significance of
$>15\sigma$,$>10\sigma$, $>5\sigma$, $>3\sigma$, and $<3\sigma$.   The region
$m_{t'}-m_N<200\,\mathrm{GeV}$ is not investigated.}
\label{fig:discoveryf}
\end{figure}

In Figure~\ref{fig:discoveryf} we plot the discovery reach as
defined by significance in (\ref{eq:disc}) for a fermionic $t'$
and scalar $N$ for $10\,\mathrm{fb}^{-1}$ of integrated luminosity
in the $m_{t'}-m_{N}$ plane.   This plot will also be
representative of the fermionic $t'$ and vector $N$ case since the
overall rate of production is independent of the spin of the $N$.
For the case of a fermionic $t'$ we find that a large region
of parameter space with only $10\;\mathrm{fb}^{-1}$ of luminosity
allows for discovery.  The maximum value for $m_{t'}$ of the
contours exhibited in Fig~\ref{fig:discoveryf} are essentially
vertical in the $m_{t'}-m_{N}$ plane as the cross section roughly
only depends on $m_{t'}$ for the bulk of the parameter space. When
$m_{t'} \approx m_t + m_N$ interference effects start to become
more important and the cross section decreases.  The region where
the interfering diagrams become important depends is somewhat
sensitive to the coupling $g_{t'N}$ but it does have any
significant impact for our study whatsoever for reasonable values
of $g_{t'N}$.

\begin{figure}[h]
\begin{center}
\includegraphics[width=8cm]{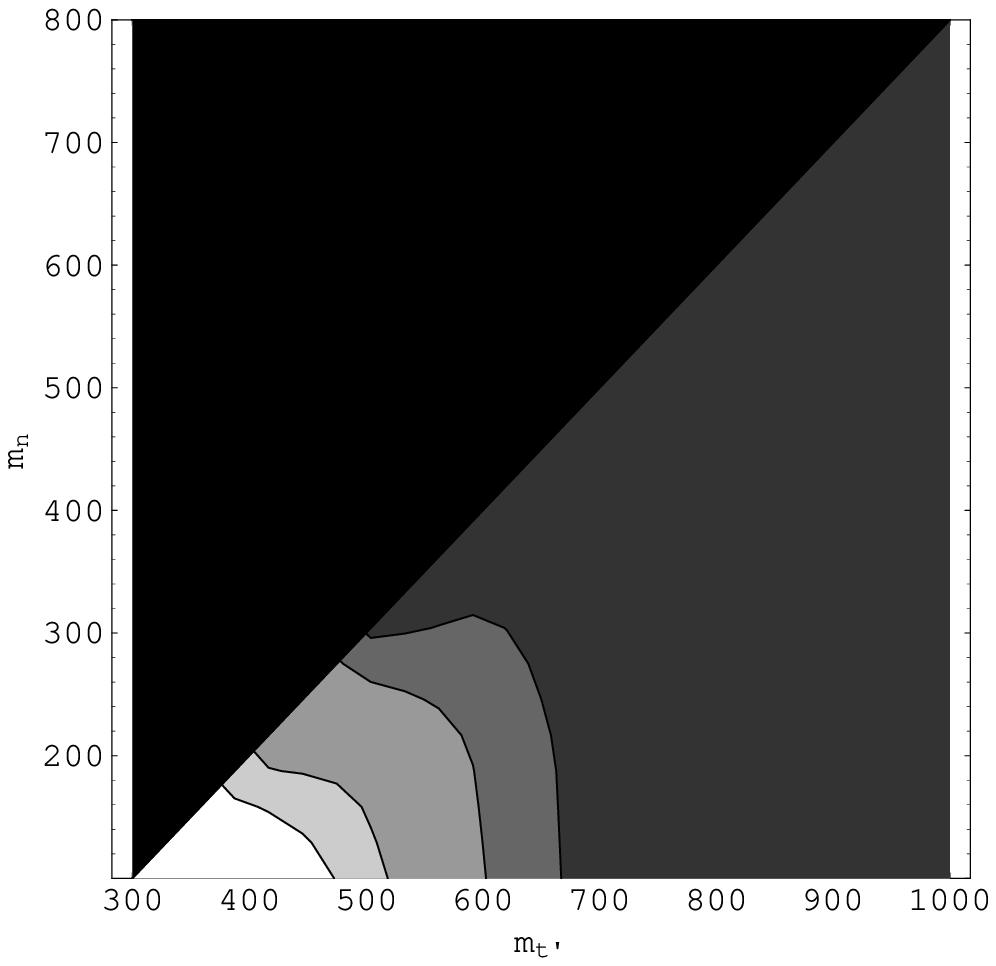}\includegraphics[width=8cm]{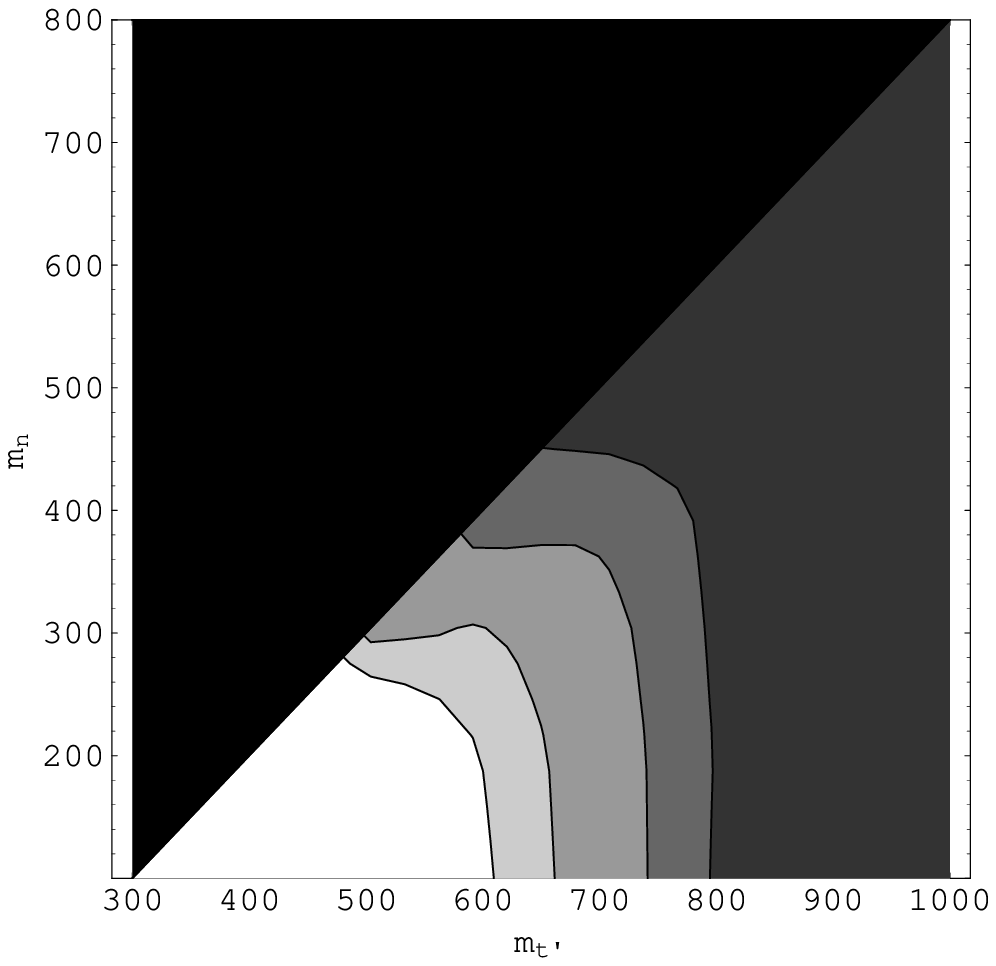}
\end{center}
\caption{Significance of signal for $t'$ scalar $N$ fermion for
$10\;\mathrm{fb}^{-1}$ on the left and and $100\,\mathrm{fb}^{-1}$
on the right. The contours (from left to right)
represent significance of
$>15\sigma$,$>10\sigma$, $>5\sigma$, $>3\sigma$, and $<3\sigma$.  The region
$m_{t'}-m_N<200\,\mathrm{GeV}$ is not investigated.}
\label{fig:discoverys}
\end{figure}

For the case of a scalar $t'$ the discovery reach is plotted for
both $10\,\mathrm{fb}^{-1}$ and $100\,\mathrm{fb}^{-1}$ in
Figure~\ref{fig:discoverys}.
For the case of a scalar $t'$ we find that with only
$10\;\mathrm{fb}^{-1}$ of luminosity the reach is only about 600 GeV
which is lower than the fermionic $t'$ case due to the smaller
cross section for scalars.  We find however that for an integrated
luminosity of $100\,\mathrm{fb}^{-1}$ the reach in the scalar $t'$
case is comparable to the fermionic case after $10\;\mathrm{fb}^{-1}$.

%
%

\begin{figure}[h]
\begin{center}
\includegraphics[width=8 cm]{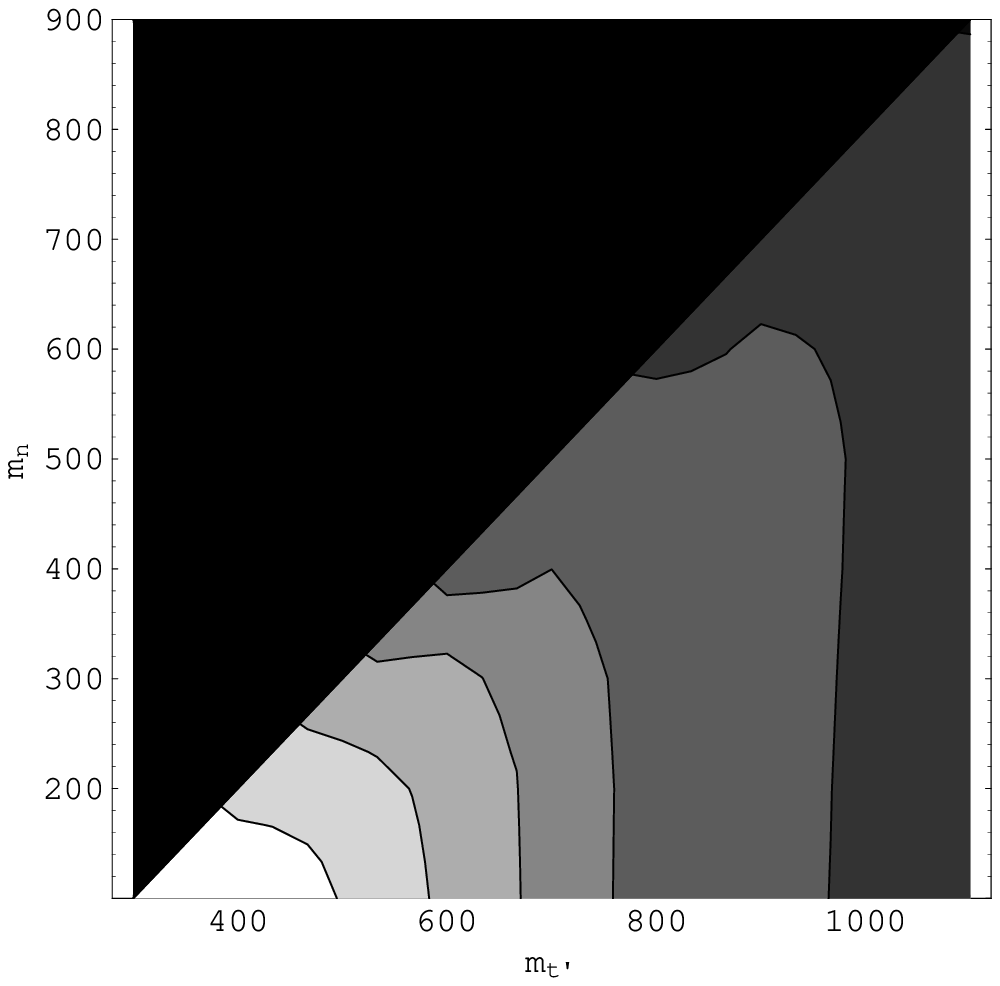}\includegraphics[width=8 cm]{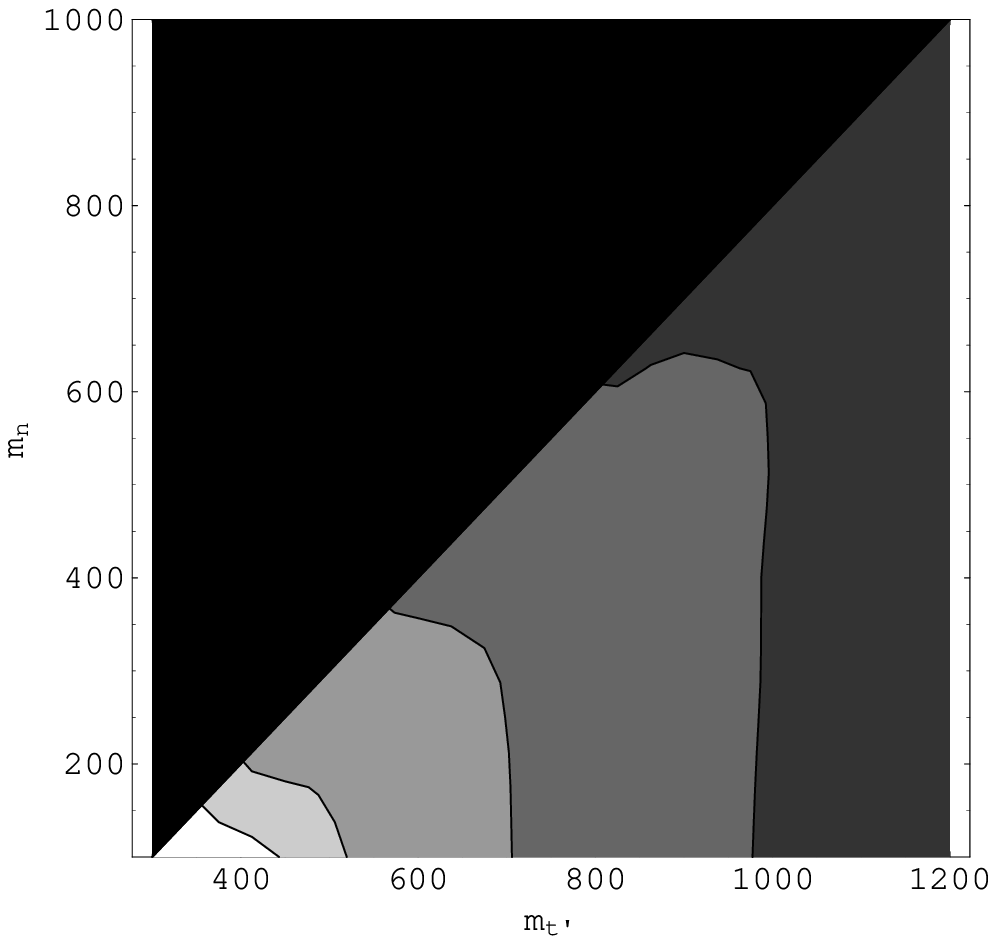}
\end{center}
\caption{Signal to background plots. At left, the case $t'$ fermion,
$N$ scalar. Contours (left to right) are $S/B = 40, 20, 10, 5, 1$.
At right, the case $t'$ scalar, $N$ fermion.
Contours are $S/B = 10, 5, 1, 0.1$.
} \label{fig:signal}
\end{figure}

While $\mathrm{signal}/\sqrt{\mathrm{background}}$ is the relevant quantity
when attempting to understand if one has discovered physics beyond
the Standard Model, the more interesting quantity for our purposes
is signal/background. We are concerned not just
with {\em discovering} new physics, but with {\em understanding} it.
For that, we would like to have a clean sample of events that we
understand to be mostly signal. This is especially important for
studying angular distributions that help to determine the $t'$ spin, as we
will discuss in Section \ref{sec:spin}. There we will be looking for
differences in the pseudorapidity distributions of top quarks.
These already small differences will be diluted if a large
number of background events are included. Having a clean sample
of signal events will make the task much easier.
We plot contours of signal to background ratio in Figure~\ref{fig:signal}.

For relatively low $t'$ masses, the sample is quite pure. For instance,
for $t'$ fermion at 300 GeV, $N$ scalar at 100 GeV, the ratio
$S/B \approx 70$. This low mass region is ideal; the cross section
is high {\em and} the signal is very pure. On the other hand,
at TeV-scale masses in the case of a fermionic $t'$ or masses
of about 700 GeV for a scalar $t'$, $S/B$ approaches 1. In that
case, it is very important to accurately understand {\em both} the
signal and background cross sections and kinematic distributions.
Given enough luminosity, at high masses the techniques we
describe in Section \ref{sec:spin} could still shed light on the
$t'$ spin, but to accurately understand the expected distributions
one needs to understand how many of the events are signal
and how many are background. For this one would ideally want
NLL calculations of expected cross sections and kinematics,
and also measurements of high-$E_T$ top quark distributions
in other channels to validate the calculations. Masses near a
TeV are a difficult challenge both theoretically and experimentally.
Still, there is a large region where $S/B$ is large enough that background
contamination is not a major worry. Our
proposed techniques could be applied in a fairly
straightforward way.

There are some foreseeable improvements of the analysis
we have outlined that could improve performance in the region
where background is neither negligible nor overwhelming.
We have described a simple set of cuts that can be applied over a wide
range of masses for the $t'$ and $N$. Once one discovers new physics
and gets a rough understanding of its mass scale, it might be
possible to develop more sophisticated cuts to keep more of
the signal relative to background. As just one example, consider
the variable $M_{T2}$ we described in Section \ref{sec:kin}. On
a sample of pure $t'\bar{t'}$ events, $M_{T2}$ will have an upper
edge $M^{max}_{T2}$. If this sample is superimposed with
background, the background events will have a smooth curve
for $M_{T2}$ lacking the upper edge. Thus a cut that
$M_{T2} < M^{max}_{T2}$ will improve the signal purity. To understand
how well this can work we would need a good detector simulation
that accurately describes the smearing of the $M_{T2}$ edge;
precisely locating this edge may be difficult.

There are other instances where altering the cuts could
help to increase the signal to background ratio. For instance,
when $m_{t'} \approx m_t + m_N$, the decay products will
generally be softer than in the typical point in parameter space.
In this case one might relax the requirement
that at least one jet have $E_T > 100$ GeV, if it is possible
to trigger on the events using some other requirement
(e.g. high $\met$ and several high-$E_T$ jets).

%
%

\section{\label{sec:massdet} Mass Determination}
\setcounter{equation}{0} \setcounter{footnote}{0}

Once we have discovered a signal in $t\bar{t}+\met$, the logical
next step is to try to determine the mass of the $t'$ and the $N$.
This is complicated by the fact that we measure the two $N$'s only
as missing transverse energy. Thus we measure $p_{x}^{N_1} +
p_{x}^{N_2}$ and $p_{y}^{N_1} + p_{y}^{N_2}$ (without information
to split the contributions of the two $N$'s) and we have no
measure of the $z$ components. On the other hand, in the top
hadronic decay mode we can attempt to reconstruct the full
four-momenta of the $t$ and $\bar{t}$, by combining jets and
demanding that each reconstruct to the top mass.

How can we use the kinematic information we have to measure
the $t'$ and $N$ masses? On an event-by-event basis,
it is apparent that we cannot. We simply do not have
enough kinematic information to pin down a
mass. On the other hand, with a large enough (and clean enough)
sample of events we can use statistical techniques to measure the
two masses. Our approach is to construct a number of variables
that each are sensitive to the masses; if they have different
contours in the $m_{t'} - m_N$ plane, then with enough of them we
can get a good estimate of the two masses.

The first kinematic variables we consider are
$\left<|\met|\right>$ and $\left<H_T\right>$, where we  average
over all events passing our cuts. In general these increase with
larger $m_{t'}$, and decrease with larger $m_N$ (as $m_N$ grows,
more of the available energy goes to its mass and less to its
$p_T$). We also consider the average $\left<M_{eff}\right>$.
Finally, we extract the kinematic edge $M_{T2}^{max}(m_{N;ref})$.

\begin{figure}[h]
\begin{center}
\includegraphics[width=6cm]{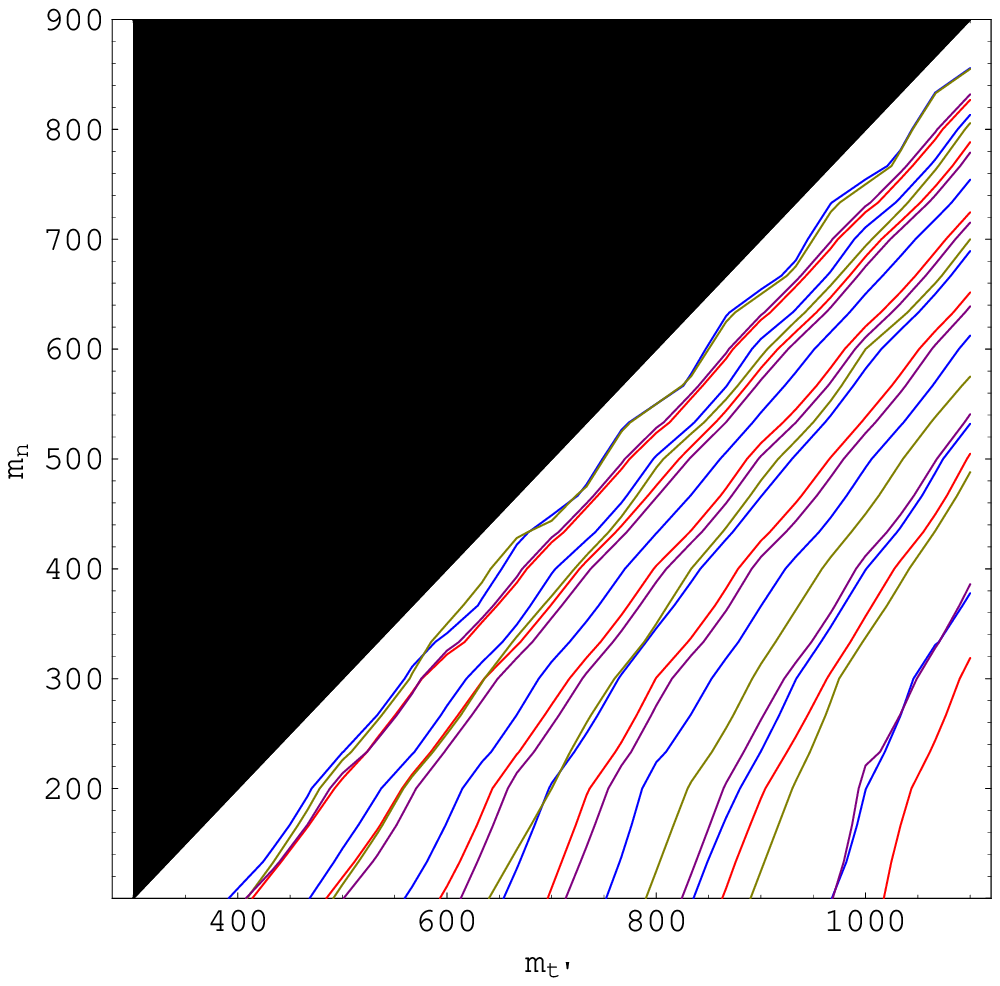}\includegraphics[width=6cm]
{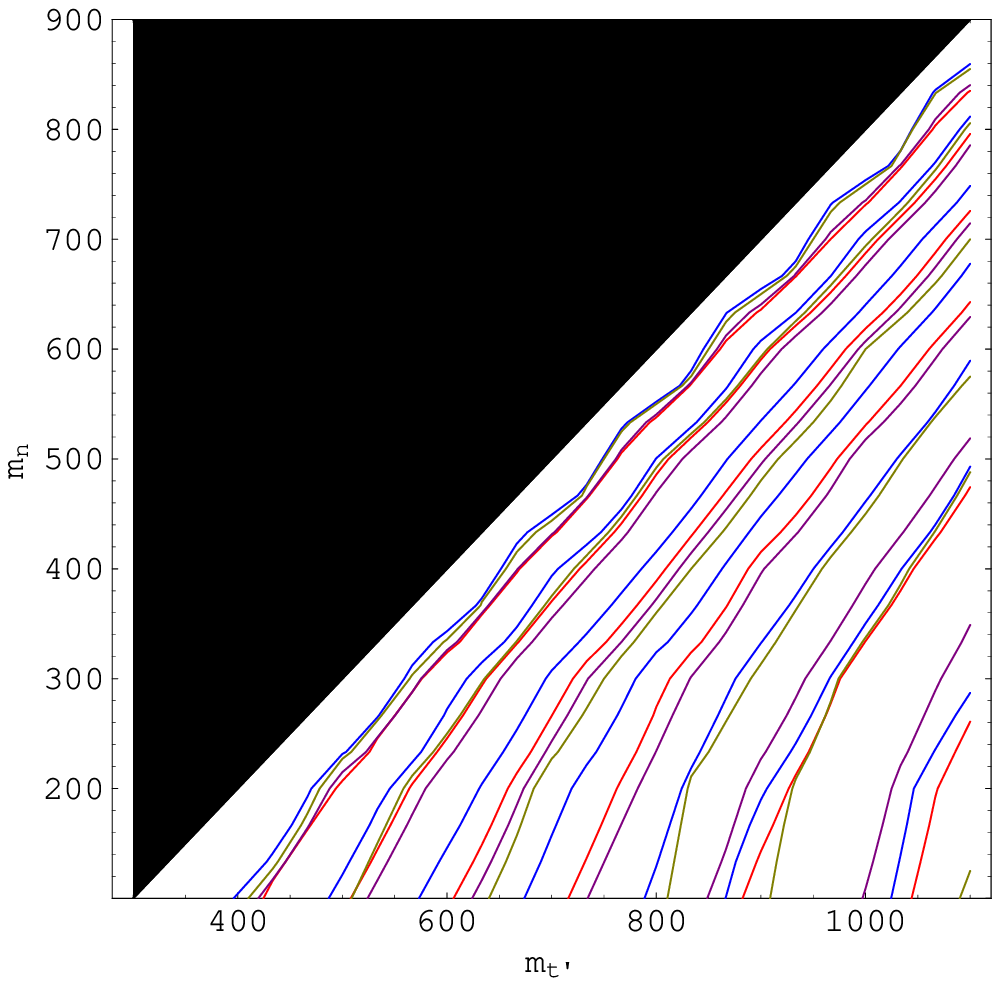}\\
\includegraphics[width=6cm]{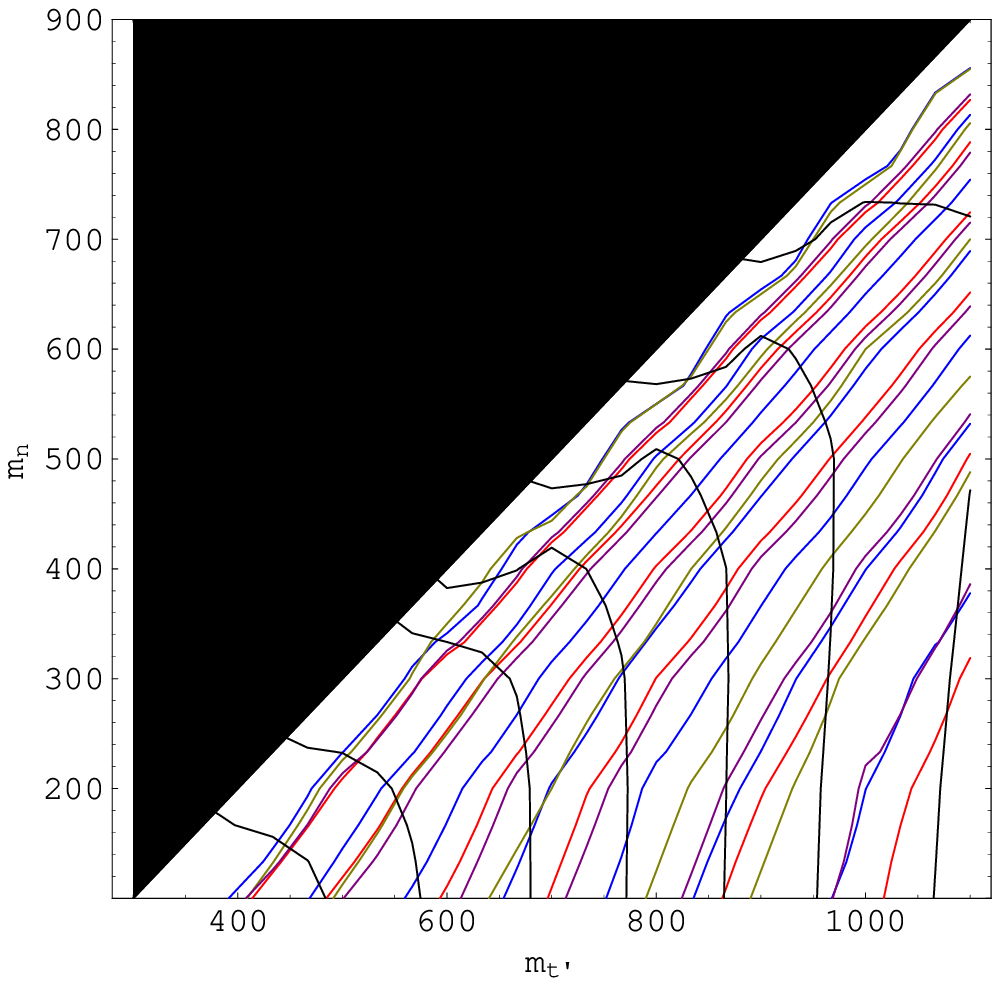}\includegraphics[width=6cm]{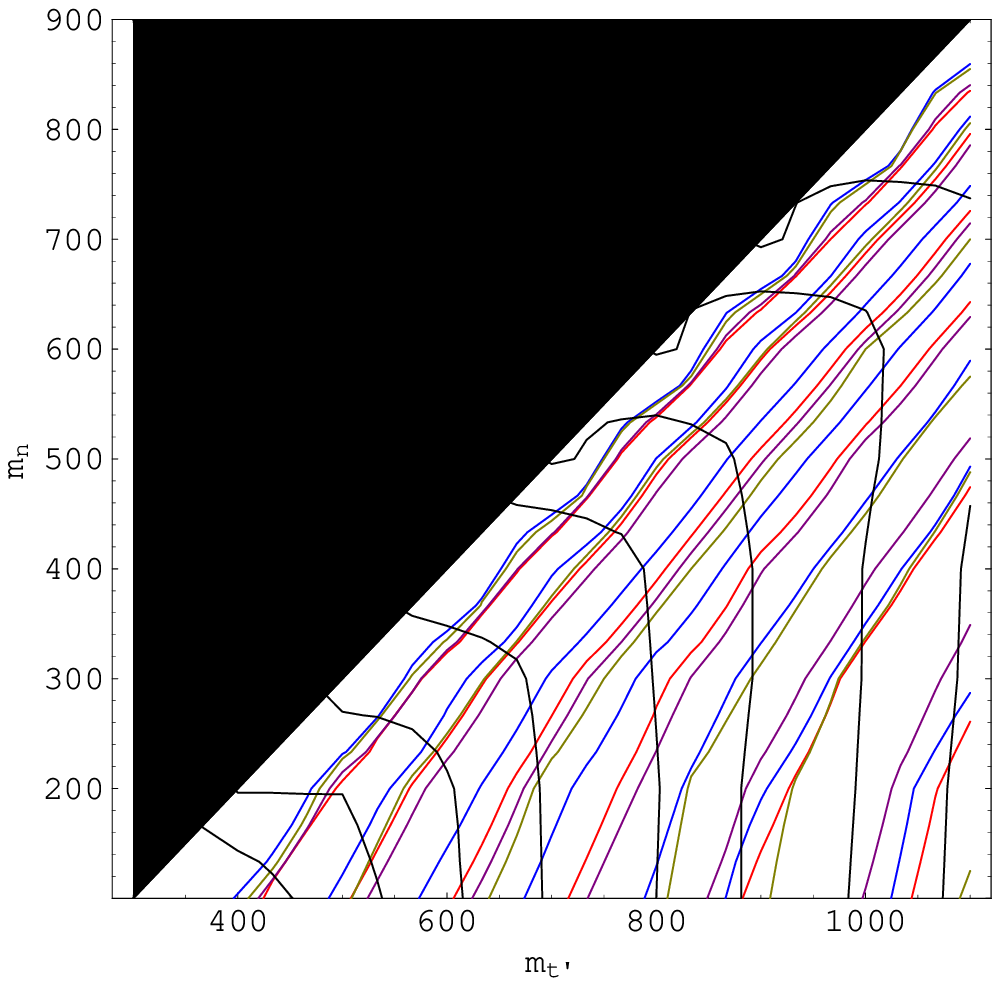}
\end{center}
\caption{Top (a): contour plots of kinematic variables,
demonstrating that they all measure the same function of
$(m_{t'},m_N)$. On the left is the case of $t'$ fermion $N$
scalar; at right, $t'$ scalar $N$ fermion. Bottom (b): the same
plots, with contours of constant cross section superimposed. At
left, $t'$ fermion $N$ scalar; at right, $t'$ scalar $N$ fermion.
Approximately, the kinematic variables are all sensitive only to
the mass {\em difference}, while the cross section is sensitive to
the $t'$ mass. ($\left<H_t\right>$ is in red,
$\left<|\met|\right>$ is in blue, $\left<M_{eff}\right>$ is in
purple, $M_{T2}^{max}$ is in gold, and in (b) $\sigma$ is in
black.)} \label{fig:contours}
\end{figure}

As it turns out, the variables $\left<|\met|\right>$,
$\left<H_T\right>$, $\left<M_{eff}\right>$, and
$M_{T2}^{max}(m_{N;ref})$ (for arbitrary $m_{N;ref}$) do not give
independent functions of $m_{t'}$ and $m_N$. This is demonstrated
clearly by a contour plot in Fig.~\ref{fig:contours}~(a). Each of
these variables is giving an approximate measurement of the mass
{\em difference} between the $t'$ and the $N$. This observation
was made independently recently in~\cite{CLW}. The noted success
of $M_{eff}$ in the mSUGRA context is due to the light neutralino
mass there; in general $M_{eff}$ does not depend strictly on the
mass of the heavy colored particle but on some function of that
mass and the mass of the LPOP.

Since any of these variables measures only roughly a mass
difference, they are insufficient to determine the two masses. On
the other hand, for a given spin the overall cross section
$\sigma$ is sensitive mostly to the mass $m_{t'}$ of the top
partner and depends strongly on $m_N$ only near the threshold for
the decay $t' \rightarrow tN$. Thus a measurement of cross section
and of another kinematic variable, say $\left<M_{eff}\right>$, is
sufficient to fix $(m_{t'}, m_N)$ {\em for a given spin of the
$t'$} as shown in Fig.~\ref{fig:contours}~(b). In general, we have
a two-fold ambiguity from this measurement, as $t'$ can be either
a fermion or a scalar. \footnote{We have checked that for $t'$ a fermion, the cross
section and kinematics are the same for $N$ either scalar or
vector. Here the coupling is fixed to be right-handed as in
Equation \ref{rhtcoup}, but these observables should be
approximately the same for the more general coupling as in
Equation \ref{mixcoup}. Preliminary checks confirm this, though
there is some deviation that must be considered in understanding
the theoretical errors on the measurement.}
We refer to these as the non-SUSY and SUSY cases
respectively. In the SUSY case, cross-sections are lower at a
given mass, so for a given measured  $(\sigma,
\left<M_{eff}\right>)$, the SUSY case corresponds to smaller
masses. We will discuss below how to exploit this. For now, we
note that given any SUSY point, there is a non-SUSY point with
matching $(\sigma, \left<M_{eff}\right>)$, obtained by raising
both the $t'$ and $N$ masses.  This translates to the fact that
any point of the plot on the right hand side of
Fig.~\ref{fig:contours}~(b) is degenerate with a point in the plot
on the left. On the other hand, the converse is not always true.
For a given non-SUSY point, if $m_N \ll m_{t'}$, there will be no
matching SUSY point. Once we reduce $m_{t'}$ to achieve the right
$\sigma$, if the splitting measured by $\left<M_{eff}\right>$ is
larger than $m_{t'}$ then we cannot find an $m_N$ to accommodate
the SUSY case. Hence, it is conceivable that a simple measurement
of cross section and effective mass can be inconsistent with SUSY.
On the other hand, if such a measurement is consistent with SUSY,
it is also consistent with the non-SUSY case and we need a new
observable to split the degeneracy.

There are some caveats to the use of the cross section $\sigma$
for mass determination, of course. For a given candidate
$(m_{t'},m_{N})$ pair and spin, one can compute the expected
$\sigma$, but this computation relies on the PDFs and on a good
understanding of backgrounds and various efficiencies.  There will
be an inherent error in the determination of the mass given that
the PDFs for gluons which will be the dominant channel at low $x$
and quarks at higher $x$ have large error bars.  For instance the
fractional error for the gluon PDF for $x$'s in the range of .1
and and less are approximately 5\% but rapidly increase for larger
$x$~\cite{PDF}.  Additionally one would need to understand the
efficiency of the triggers relevant to this process since we are
relying upon being able to really understand the actual number of
events after reconstruction. The various efficiencies that are
important for our study however should be better pinned down once
a more complete understanding of the detectors is accomplished;
however at most the result of this understanding will be to simply
scale our cross sections by a number less than one but presumably
very close to one given our conservative estimates.  Despite the
various concerns over the use of cross section, it is important to
keep in mind that the dependence of the cross section on the $t'$
mass is very strong so a large uncertainty on the cross section
translates into a much smaller uncertainty on the $t'$ mass.  If
there were a 100\% error in the cross section it would be
reflected as approximately a 100 GeV error in the bulk of our
parameter space. For this reason we believe that even with the
inherent uncertainties in the cross section measurement it can
still be used for a reliable estimate of the masses.

The degeneracy we have pointed out, even after using cross section
information to determine masses up to the choice of spin, is often
not taken into account in recent studies attempting to distinguish
SUSY and non-SUSY cases where the starting point is the same
spectrum instead of the same cross sections and kinematic
observables~\cite{Barr,SmillieWebber,IsItSUSY,NewBarr}. This is certainly a
conceivable starting point given that one may be able to measure
the spectrum independently in enough other channels, however from
a bottom up perspective the degeneracy we demonstrate here seems
more likely to be a concern if one is not in such an optimistic
region as examined in other studies.  It has been conjectured that
with enough channels the ability to distinguish SUSY from a
non-SUSY case is possible when looking at patterns of signatures,
while any given channel can be made degenerate~\cite{IsItSUSY}.
However, within the MSSM alone there are many degeneracies amongst
parameters when looking at a large set of inclusive
signatures~\cite{lotsadegeneracies}, thus it is quite plausible to
believe these degeneracies are no less frequent in the SUSY vs.
non-SUSY cases especially given that new physics may not show up
in multiple channels.

\section{Spin Determination}
\label{sec:spin}
\setcounter{equation}{0} \setcounter{footnote}{0}

We have seen that kinematic variables, together with cross
section, are sufficient to give estimates of the masses $m_{t'}$,
$m_N$ for a {\em given choice} of spins, but that in general they
do not distinguish the supersymmetric case from the non-SUSY case.
To distinguish the various cases and also find the correct masses
one has to find a method to distinguish the spins.  There have
been some attempts at coming up with techniques for spin
determination at the
LHC~\cite{Barr,SmillieWebber,IsItSUSY,NewBarr}, however as
mentioned before the starting point of these studies has been to
choose an identical mass spectrum for a SUSY vs non-SUSY case.
These studies do not break the degeneracy that we are interested
in and thus do not distinguish the spins in the most general case.
It is important to note though that given equivalent kinematics
and cross section, the mass scale of the SUSY case is much lower.
This is useful: we do not have to measure just spin correlations
to distinguish the two cases. Instead, we can look for other
quantities that distinguish the overall mass scale.\footnote{Of
course, one might obtain information from other sources. For
instance, determination of the $N$ mass in a dark matter
direct-detection experiment would allow us to obtain $m_{t'}$ from
$M^{max}_{T2}(m_N)$, and then we could simply read off the $t'$
spin from the cross section as suggested in
\cite{IsItSUSY,BattagliaCLIC,SmillieWebber}. However, we will be
pessimistic and assume we must learn everything from the
$t\bar{t}+\met$ signal.}

In general, spin correlations are studied by boosting to some
optimal frame, choosing a plane, and then constructing some
asymmetry from the distribution of decay products from the two
original particles about the chosen plane. However, since we
cannot resolve the direction of the two $N$ particles, it is
difficult to boost to a useful frame. The simplest thing that we
can do is to try to build asymmetries in the lab frame, based on
the momenta of the tops.

We define two candidate asymmetries:
\begin{itemize}
\item The {\bf beam-line asymmetry} (BLA): Let $p^{t1}_z$ and
$p^{t2}_z$ be the components of the top momenta along the
beam-axis, in the lab frame. Define $N^z_+$ to be the number of
events where $p^{t1}_z p^{t2}_z > 0$ and $N^z_-$ to be the number
of events where $p^{t1}_z p^{t2}_z < 0$. Then our asymmetry is
$BLA = (N^z_+ - N^z_-)/(N^z_+ + N^z_-)$. Equivalently, BLA can
be defined based on the pseudorapidities of the tops, which have
the same sign as $p_z$.

\item The {\bf
directional asymmetry} (DA): Consider $\cos \theta_{t\bar{t}}$
where $\theta_{t\bar{t}}$ is the angle between the top momenta in
the lab frame. In general the cross section will have a linear
dependence on $\cos \theta_{t\bar{t}}$, but isolation cuts remove
the events where this is near 1. We define $N^d_+$ as the number
of events where $0.5 < \cos \theta_{t\bar{t}} < 0.9$ and $N^d_-$
as the number of events where $-0.9 < \cos \theta_{t\bar{t}} <
-0.5$. Our asymmetry is $DA = (N^d_+ - N^d_-)/(N^d_+ + N^d_-)$.
\end{itemize}

These asymmetries convolve effects of spin correlations and of
overall mass scale. We have listed them for a variety of points
with similar kinematics and cross section in Table \ref{tab:asym}.
They are both consistently larger in the SUSY case. For the BLA,
it is particularly easy to understand how the mass scale affects
the quantity we compute, since the BLA can be defined using
only pseudorapidities. First consider the
SUSY case. We produce two scalar tops, so there can be no initial
spin correlation. In the center-of-mass frame of the colliding
partons, the stops are back-to-back. Each stop decays
isotropically in its rest frame. Thus, in the center-of-mass frame
of the colliding partons, the two tops have a higher probability
of moving in opposite directions with respect to a given plane
than of moving in the same direction. In the case of a fermionic
$t'$, the expected asymmetry is modified, because the spins of the
initial $t'\bar{t'}$ pair are correlated. In either case, the
center-of-mass frame differs from the lab frame by some overall
boost along the beam axis. The parton distribution functions are
rapidly falling functions of $x$, so at heavier masses one should
expect a smaller average boost. On the other hand at lower masses
the boost will often be large enough that the BLA becomes
significantly greater than zero. In this way the BLA will give
some indication of the $x$ values of the colliding partons. This is clear
in Table \ref{tab:asym}; for a given spin, the asymmetry falls off
at higher masses.

\begin{figure}[h]
\begin{center}
\includegraphics[width=8cm]{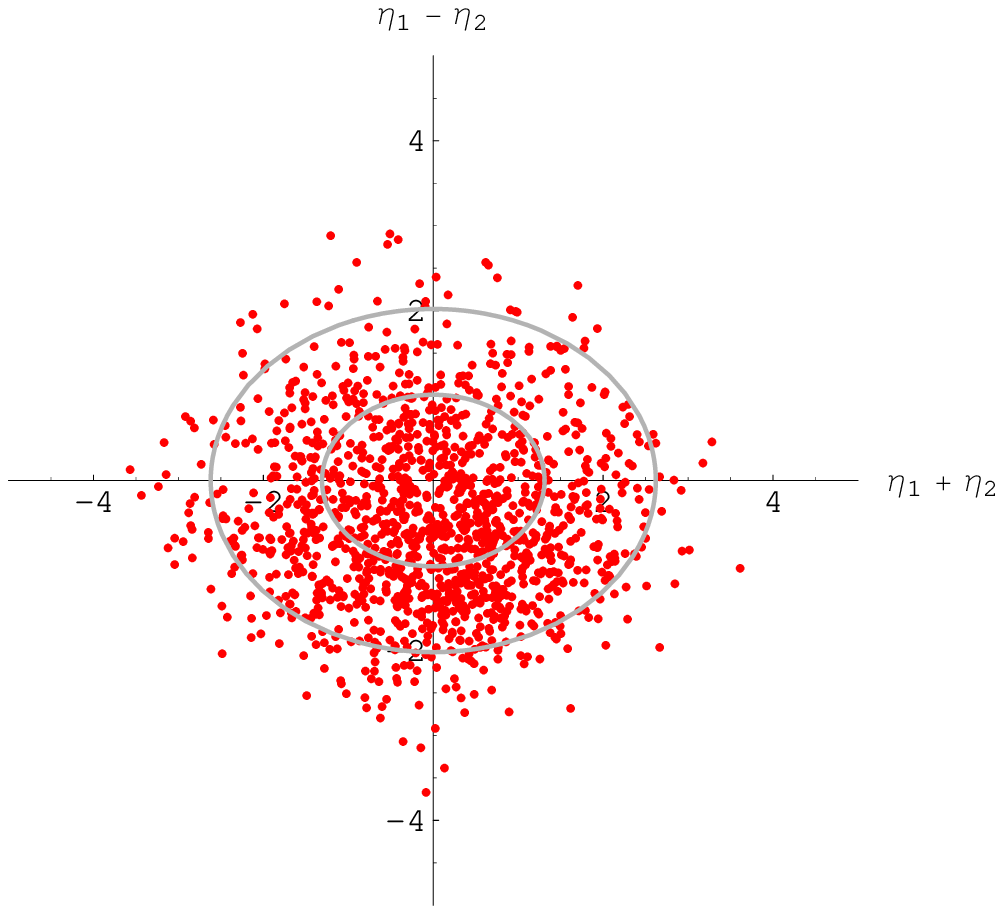}\includegraphics[width=8cm]{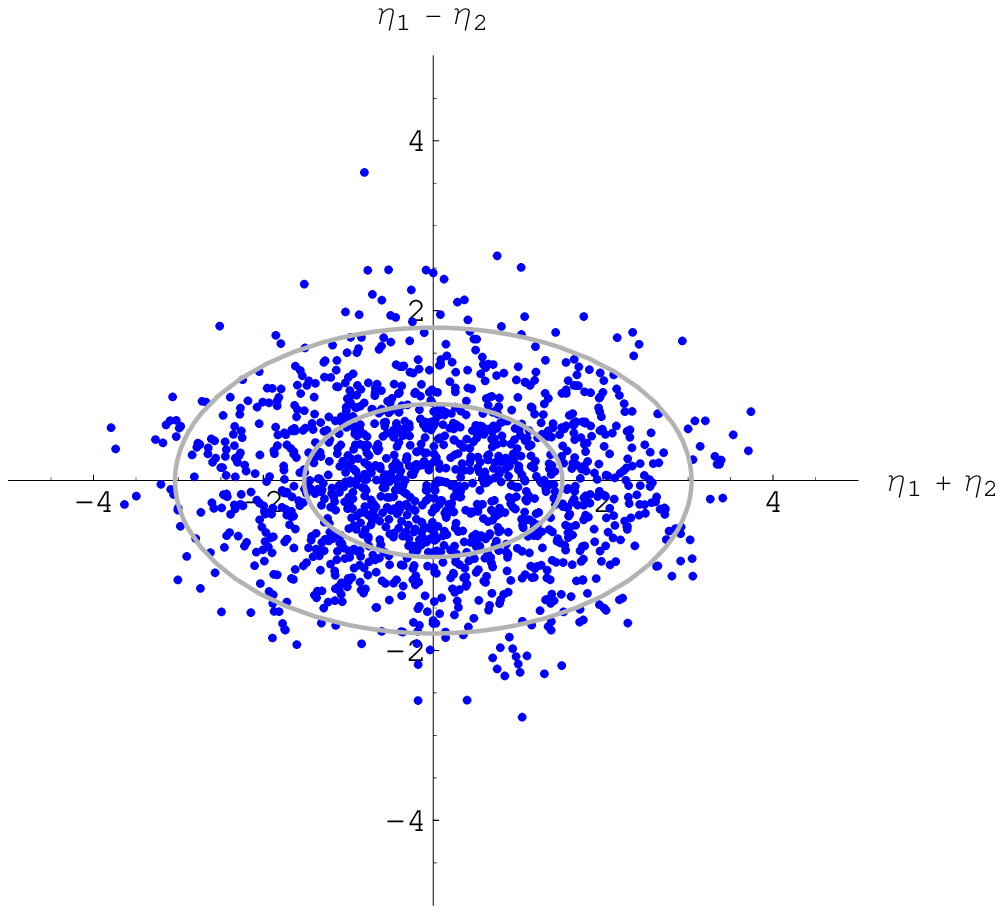}
\end{center}
\caption{Distribution of events in the $(\eta_+,\eta_-)$ plane for two points
with similar cross-section and kinematics but different spins, with one and two
sigma contours. At left: $t'$ fermion,
mass 700 GeV, $N$ scalar, mass 400 GeV, $(\sigma_+, \sigma_-) = (1.31,1.01)$; at right, $t'$ scalar, mass 500 GeV,
$N$ fermion, mass 150 GeV, $(\sigma_+, \sigma_-) = (1.52,0.90)$. In the lighter ($t'$ scalar) case, there is
on average more boost, so the ellipse is stretched more along the $\eta_+$ axis.}
\label{fig:etaplots}
\end{figure}

In addition to these asymmetries, there is a related technique for
studying the effect of the mass scale on the observed data. We
have noted that the BLA can be defined in terms of pseudorapidity.
It is instructive to study the pseudorapidity distributions of the
two tops directly. The difference $\eta_- = \eta_1 - \eta_2$ is
invariant under boosts along the beam axis, but the sum $\eta_+ =
\eta_1 + \eta_2$ is not. We plot events for the different spin
cases in the $(\eta_+,\eta_-)$ plane in the figure
\ref{fig:etaplots}, for a pair of points with matching cross
section and kinematics. It appears that to a good approximation
the events are Gaussian distributed in the two variables. In this
way the distributions define an ellipse; in the scalar case, this
ellipse is stretched out significantly more on the $\eta_+$ axis.
The observation that the scalar case is more stretched out when
looking at a particular degeneracy is obviously commensurate with
the results of the BLA and DA.  The additional use of the
information related to $\eta_-$ has been argued to be sensitive
only to spin correlation~\cite{NewBarr} in certain cascade decays.
This again was studied in the context of a degenerate mass
spectrum thus avoiding the degeneracies associated with different
mass scales.  When investigating the degeneracies we presented in
Section~\ref{sec:massdet} it is almost impossible to entirely
deconvolve the effects of mass scale and spin correlation. That is
why we advocate the asymmetries as well as the full correlation in
pseudorapidity where both the $\eta_+$ and $\eta_-$ information
are useful, and depending on the masses involved one may have a
much stronger effect at distinguishing the spins.

To quantify the results of pseudorapidity correlations, we perform
a likelihood fit for the standard deviations $\sigma_+$ and
$\sigma_-$, maximizing $-2\log{\cal L}$ where ${\cal L} =
\prod_{i} P(\eta_{+i}, \eta_{-i})$ with
\begin{equation}
P(\eta_+,\eta_-) = \frac{C}{\sigma_+ \sigma_-} \exp\left(-
\frac{\eta_+^2}{2\sigma_+^2}-\frac{\eta_-^2}{2\sigma_-^2}\right).
\end{equation}
The likelihood fit was performed using Minuit~\cite{Minuit}.
Minuit returns errors on the fitted parameters $\sigma_+$ and
$\sigma_-$. These errors should scale like $1/\sqrt{N_{events}}$,
and this is consistent with the samples of unweighted events we
have analyzed. The coefficient, as reported by Minuit, is
approximately 1.\footnote{Obtaining large numbers of {\em
unweighted} events to systematically explore the error on various
random subsamples is computationally intensive, but on a variety
of samples up to 7,000 events we obtain error estimates ranging
from $0.93/\sqrt{N_{events}}$ to $1.04/\sqrt{N_{events}}$.} We
view the Minuit errors as approximately describing the
experimental error after a given number of events, though other
effects must be considered. (For instance, finite $\eta$
resolution will have some effect, although if the smearing of the
$\eta$'s of the two tops is uncorrelated it does not seem that it
should pose major problems for our fit.)

As an example, consider the degeneracy of $t'$ fermion at 700 GeV
and $N$ scalar at 400 GeV with $t'$ scalar at 500 GeV and $N$
fermion at 150 GeV. The corresponding fitted values of $\sigma_+$
are 1.31 and 1.52 respectively. The cross section (with all
branching ratios and efficiencies taken into account) is about 2.0
fb. Ignoring background (which is a factor of 5 smaller), one will
have about 600 signal events after 300 fb$^{-1}$, and
$1/\sqrt{N_{events}}$ is $\approx 0.04$. Taking this to be a good
measure of the error, at this point the fitted values are
separated by 5~$\sigma$. With about 1200 fb$^{-1}$ of luminosity,
the fitted values are separated by 10~$\sigma$. Here we are
restricting ourselves to considering the right-handed coupling Eq.
\ref{rhtcoup}. If we allow a more general Lorentz structure, there
might be some small range around these fitted values allowed for a
given choice of $t'$ spin, so one will want to pick the most
pessimistic separation, and the luminosity required might be
somewhat larger. (Unless some additional measurement can help fix
the Lorentz structure of the coupling.) Similar considerations
will apply to BLA and DA, and to $\sigma_-$, which grows smaller
in the scalar case. One can measure each of these quantities for
more consistency, but they are highly correlated, so they do not
give independent checks of the $t'$ spin.

Of course, how confidently one can say a measured set of data
corresponds to one case or the other will depend on the measured
central value. Given the possibility that mismeasurement and other
considerations will lead to the actual efficiency being somewhat
less than our estimates, we think it best to be cautious rather
than make a definitive claim about the luminosity needed to
distinguish spins. Ultimately a detailed study with detector
simulation, larger samples, and inclusion of background together
with signal is necessary to understand exactly the scope of
applicability of this technique, but it appears promising. It
appears that with several hundred fb$^{-1}$ up to a few thousand
fb$^{-1}$ one can use angular observables to resolve the spin of
the $t'$ over a wide range of masses. Making this discrimination
could require the LHC luminosity upgrade \cite{LHCupgrade}.  It is
interesting to note that the LHC may be more feasible for spin
determination than a linear collider dependent on the values of
$m_{t'}$; future study of this is worth investigating.

\begin{table}[ht]
\centering
\begin{tabular}{llllllll}
Spin ($t',N$) & $(m_{t'},m_N)$ & $\left<H_t\right>$ & $\sigma$ &
BLA
& DA & $\sigma_{+}$ & $\sigma_-$\\
\hline
(F,S) & (550,300) & 781 & 5.1 &  0.22 & -0.43 & 1.40 & 1.05 \\
(S,F) & (390,115) & 786 & 5.0 & 0.31 & -0.25 & 1.59 & 0.94 \\
(F,V) & (550,300) & 779 & 5.2 & 0.22 & -0.46 & 1.39 & 1.03 \\
\hline
(F,S) & (600,350) & 775 & 3.3 & 0.16 & -0.44 & 1.38 & 1.15 \\
(S,F) & (415,165) & 777 & 3.1 & 0.32 & -0.34 & 1.57 & 0.82 \\
(F,V) & (600,350) & 785 & 3.4 & 0.20 & -0.46 & 1.37 & 1.00 \\
\hline
(F,S) & (650,350) & 860 & 3.1 & 0.16 & -0.41 & 1.35 & 0.99 \\
(S,F) & (475,100) & 863 & 2.9 & 0.30 & -0.23 & 1.53 & 0.92 \\
(F,V) & (650,350) & 860 & 3.2 &  0.19 & -0.40 & 1.34 & 1.05\\
\hline
(F,S) & (700,400) & 865 & 2.0 & 0.16 & -0.40 & 1.31 & 1.01 \\
(S,F) & (500,150) & 874 & 2.1 & 0.26 & -0.32 & 1.52 & 0.90\\
(F,V) & (700,400) & 857 & 2.1 & 0.16 & -0.45 & 1.30 & 1.08 \\
\hline
(F,S) & (700,500) & 695 & 0.51 & 0.19 & -0.66 & 1.27 & 1.03 \\
(S,F) & (515,315) & 742 & 0.44 & 0.36 & -0.55 &  1.40 & 0.75 \\
(F,V) & (700,500) & 690 & 0.50 & 0.17 & -0.64 & 1.20 & 0.94 \\
\hline
(F,S) & (750,425) & 904 & 1.6 & 0.15 & -0.39 & 1.32 & 1.00 \\
(S,F) & (550,150) & 917 & 1.5 & 0.21 & -0.23 & 1.51 & 0.93 \\
(F,V) & (750,425) & 896 & 1.6 & 0.14 & -0.37 & 1.30 & 1.05 \\
\hline
(F,S) & (800,450) & 943 & 1.2 & 0.13 & -0.34 &  1.28 & 1.07 \\
(S,F) & (575,125) & 956 & 1.2 & 0.24 & -0.24 & 1.45 & 0.97 \\
(F,V) & (800,450) & 936 & 1.2 & 0.13 & -0.34 & 1.30 & 1.03 \\
\hline
(F,S) & (900,500) & 1019 & 0.66 & 0.087 & -0.29 & 1.25 & 1.11 \\
(S,F) & (645,60) & 1043 & 0.68 & 0.15 & -0.22 &  1.45 & 1.08 \\
(F,V) & (900,500) & 1012 & 0.66 & 0.12 & -0.32 & 1.28 & 1.11 \\
\hline
(F,S) & (900,550) & 953 & 0.59 & 0.11 & -0.36 & 1.23 & 1.07  \\
(S,F) & (645,235) & 967 & 0.61 & 0.20 & -0.26 & 1.41 & 1.04 \\
(F,V) & (900,550) & 947 & 0.60 & 0.13 & -0.38 & 1.26 & 1.07 \\
\hline
\end{tabular}
\caption{
Asymmetries and rapidity ellipse shapes for points with matching kinematics. Masses and $H_t$ in GeV, $\sigma$ in fb. S denotes spin 0, F denotes spin $1/2$, V denotes spin 1. These numbers are reported
for {\em signal only}; for the points with lower cross section, one needs to consider the effect of the
background on the measured quantities. Signal becomes diluted at high masses, so discriminating
among the spins grows more difficult.}
\label{tab:asym}
\end{table}

\section{Relevance To General Scenarios}
\label{sec:nonSMbg} \setcounter{equation}{0}
\setcounter{footnote}{0}

We have focused exclusively on the signature $t\bar{t}+\met$. In
most models of new physics there will be additional new physics
states accessible at the LHC. It is important to ask whether our
discussion is still applicable in such cases. Signatures from
other new states will, of course, provide other clues to the
underlying model, but they also could potentially make the
analysis we have outlined more difficult. There are multiple
sources of complication: new backgrounds, new decay modes, and new
production channels. Our method should be applicable in a wide
variety of cases, despite the added complications.

In a model with additional parity-odd states, there are
potentially new backgrounds to our signal. The most obvious is an
additional (somewhat heavier) parity-odd top. In SUSY, for
instance, one always expects two scalar tops, from the mixing of
the partners of $t_L$ and $t_R$. This is clearly a background that
we cannot eliminate by harder kinematic cuts. However, it should
be relatively straightforward to deal with in many scenarios.
If the additional
parity-odd top is so heavy that its cross section is well below
that of the lightest $t'$, we can ignore it when performing our
analysis. If it is light enough that it has a significant cross
section (well above SM backgrounds), then it would show up on the
$M_{T2}$ plot as a second edge. By the relative location and height
of the two edges, one could possibly learn the difference in mass and in cross
section of the two parity-odd tops. This is a case where $M_{T2}$
could give us clear new information, and deserves further study.
The most troublesome case would be very nearly degenerate
partners of the $t_L$ and $t_R$, since they would be nearly
indistinguishable but would increase the overall cross section.
This requires further study.

As another example involving new parity-odd states, a sufficiently
heavy parity-odd $Z'$ might decay to $t'$ + $\bar{t}$, for
instance. Production of $Z'N$, then, would give $t\bar{t} + \met$.
This would be quite difficult to distinguish from our signal, but
the cross section should be smaller due to the the weak couplings
involved in the production, and there is some branching fraction
$Br(Z' \rightarrow t'\bar{t})$ multiplying this cross section as
well.

New parity-even states can also give backgrounds to our signal.
For instance, in the Littlest Higgs with T-parity one can have a T-even
top $t'_+$ that has some branching fraction to $t' N$. Single
production of $t'_+$ is the dominant channel, but it is generally
produced in association with a light-quark jet. Pair production of
$t'_+$ would fake our signal, and could happen in the low-mass
region. If the branching fraction and cross section are large
enough that this fakes our signal at a high rate (implausible in
the Littlest Higgs with T-parity, but possible since we are
thinking model-independently), one should consider the use of
$M_{T2}$ to find an additional edge as we discussed for the extra
odd $t'$.

Less directly, long decay chains could provide events with
multiple jets and $\met$. As with Standard Model backgrounds, if
actual tops are not produced, cuts on mass reconstruction and
$b$-tagging will remove much of such backgrounds. However, some
backgrounds could be more troublesome. For instance, in SUSY one
could have $\tilde{g}\tilde{q}$ with $\tilde{g} \rightarrow
\tilde{t}t$. Vetoing additional hard jets could be useful in such
events, but one needs a good understanding of pile-up, initial
state radiation, and other issues that could lead to additional
jets in the signal itself. Whether these cuts will be sufficient
for a clean $t'\bar{t'}$ sample will depend on the details of the
mass spectrum under consideration, but it is plausible that we can
reject most of the backgrounds without sacrificing too much
efficiency on the signal. On the other hand, an ATLAS study of a
particular mSUGRA point found that it was difficult to obtain a
very pure sample of $\tilde{t}\bar{\tilde{t}}$ production over
SUSY backgrounds \cite{ATDR,ATLASstop}. Again, this is a case where
intelligent use of $M_{T2}$ might provide useful new information,
and much will depend on the details of the spectrum.

Another  consideration is that in a more complicated model, the
$t'$ is likely to have other decay channels. For instance, one
would might to consider $W'b$, $Wb'$, $Z't$, or $Z'c$. In general
these will have similar signatures to our $tN$ decay: for
instance, $W'b \rightarrow jjbN$, but the $jjb$ mass will no
longer be constrained to equal the top mass. To measure the
branching fractions one wants to try to find each of these modes,
but the lack of mass window cuts implies that they are more
susceptible to multi-jet SM backgrounds. This deserves further
study. If all the decay channels can be measured reasonably well,
one can estimate the branching fraction to $tN$ and apply our
analysis. So long as the $t' \rightarrow tN$ branching fraction is
${\cal O}(1)$, and new physics backgrounds do not swamp the
signal, one should still be able to apply our analysis. (Of
course, more luminosity might be needed.)

A further consideration in the presence of other new physics is
that new production channels might open up for the $t'$. For
instance, a relatively light gluino in SUSY would alter the
production cross section. Thus, to apply our analysis of mass
determination using cross section, one would need some idea of how
much the cross section is altered by the other new physics.
However, as we have noted, even large uncertainties in the cross
section translate to relatively small uncertainties in masses. It
seems unlikely that new physics will cause large differences in
cross section, so this might not be very problematic.

There is much more work to be done to understand how further
details of the model can complicate an analysis of the sort we
have outlined. However, it appears that in a large variety of
models $t' \rightarrow tN$ is a significant decay mode that can be
observed cleanly. Once new physics is observed in several
channels, one can start to understand how the uncertainties in our
analysis are affected. The claim that at the same mass scale
scalar top cross sections are significantly smaller than fermionic
$t'$ cross sections is a robust one, and we expect that the LHC
can discriminate the two spins in a wide variety of models. It is
also worth pointing out that new physics might not be described
by any current model. It is conceivable that new physics will
have a relatively light $t'$ with no other states light enough to affect
our results.

\section{Conclusions}
\setcounter{equation}{0}
\setcounter{footnote}{0}

In this paper we have put forth a new perspective for studying the
phenomenology of physics beyond the SM that directly relates to
the LHC. We have advocated a model independent signature based
analysis motivated by naturalness to develop new tools, and
we have hopefully dispelled certain misconceptions. We have begun
utilizing this new methodology by analyzing the first new particle
dictated by naturalness, the partner of the SM top quark, $t'$.

Based on the model independent framework that we set up in this
paper of having a parity odd $t'$ decay to a right handed top
quark and the lightest parity odd particle $N$ we set out to
answer the following questions:
\begin{itemize}
\item Can the signature be observed at high significance over the
SM backgrounds? \item How well can we determine the masses of the
$t'$ and $N$? \item Can we devise an algorithm for measuring the
spin of the $t'$ or the $N$?
\end{itemize}
We have shown that in the hadronic channel for a large range of
$t'$ mass the $t'$ has a large significance and can be discovered
at the LHC within a short running time.  We additionally
demonstrated over a large range of $t'$ mass the ratio of signal
to background is greater than 1.  We then set up a program for
measuring the masses of $t'$ and $N$.  We showed that standard
kinematic observables measure only one independent function of
$m_{t'}$ and $m_N$, which is approximately the mass {\em
difference} over much of parameter space.  Additionally we
demonstrated that certain kinematic observables that have been
used in the past to determine the mass of strongly interacting
particles are only justified in certain regions.  The way we found
most effective for measuring both the mass of $t'$ and $N$ was to
combine any standard kinematic observable with the cross section.
We found though that this \emph{only} determines the masses up to
a discrete choice of spin. Given a scalar $t'$ (SUSY case) there
always is a corresponding fermionic $t'$ (non SUSY) with a heavier
mass that has the same kinematic observables. This degeneracy is
not accounted for in many studies comparing models where the
starting point is often to choose the same mass spectrum.  To
break the degeneracy we needed to determine spin information from
the LHC. We introduced several possible ways to break the spin
degeneracy. Most promising was using the additional kinematic
information about the rapidities of the top quark in the hadronic
channel.  We put forth a new asymmetry as well as studying
correlations amongst rapidities that can be used to determine the
spin of $t'$, and thus pin down the masses of the particles.  This
can be taken heuristically as a way to determine the difference
between SUSY and and other models.

There are several logical extensions to our work.  In focusing on
the top sector we have narrowed the scope of our study to
discovery possibilities and determination of basic properties such
as mass and spin.  If naturalness is really playing a role in TeV
scale physics one would need to test this by determining the
couplings of the $t'$ to the Higgs. One could also extend our
study to an even more generic coupling to the top quark as well as
including a partner of the bottom quark. From this bottom up model
independent point of view a logical next step would also be to add
in more particles dictated by naturalness, i.e. partners of the SM
gauge bosons and Higgs.  We leave these possibilities as well as
others that can be motivated from the philosophy set forth in this
paper for future study.

\section*{Acknowledgments}

We thank Ulrich Heintz, Meenakshi Narain, Peter Onyisi,
Maxim Perelstein, and Matt Strassler for useful
conversations. We thank Adam Aurisano and Seth Zenz for directing
us to useful references on related experimental issues. We thank
Fabio Maltoni and Tim Stelzer for answering MadGraph questions. PM
would like to thank the theory groups at Boston University, Harvard University,
and the University of Washington for their hospitality
while this paper was in preparation. PM and
MR were supported in part by the National Science Foundation under
grant PHY-0355005 and MR was supported by a National Science
Foundation Graduate Research Fellowship.


\begin{thebibliography}{99}

\bibitem{LHoriginal}
  N.~Arkani-Hamed, A.~G.~Cohen, E.~Katz and A.~E.~Nelson,
  JHEP {\bf 0207}, 034 (2002)
  [arXiv:hep-ph/0206021].


\bibitem{Rparity}
  G.~R.~Farrar and P.~Fayet,
  Phys.\ Lett.\ B {\bf 76}, 575 (1978).

\bibitem{MParity}
S.~Dimopoulos and H.~Georgi,
  Nucl.\ Phys.\ B {\bf 193}, 150 (1981).
  N.~Sakai and T.~Yanagida,
  Nucl.\ Phys.\ B {\bf 197}, 533 (1982).
   S.~Dimopoulos, S.~Raby and F.~Wilczek,
  Phys.\ Lett.\ B {\bf 112} (1982) 133.

\bibitem{TParity}
  H.~C.~Cheng and I.~Low,
  JHEP {\bf 0309}, 051 (2003)
  [arXiv:hep-ph/0308199],
  H.~C.~Cheng and I.~Low,
  JHEP {\bf 0408}, 061 (2004)
  [arXiv:hep-ph/0405243],
  I.~Low,
  JHEP {\bf 0410}, 067 (2004)
  [arXiv:hep-ph/0409025].


\bibitem{UED}
  T.~Appelquist, H.~C.~Cheng and B.~A.~Dobrescu,
  Phys.\ Rev.\ D {\bf 64}, 035002 (2001)
  [arXiv:hep-ph/0012100].

\bibitem{BosonicSUSY}
  H.~C.~Cheng, K.~T.~Matchev and M.~Schmaltz,
  Phys.\ Rev.\ D {\bf 66}, 056006 (2002)
  [arXiv:hep-ph/0205314].


 \bibitem{jaypatrick}
  J.~Hubisz and P.~Meade,
  Phys.\ Rev.\ D {\bf 71}, 035016 (2005)
  [arXiv:hep-ph/0411264].

\bibitem{CLW}
  H.~C.~Cheng, I.~Low and L.~T.~Wang,
  arXiv:hep-ph/0510225.


\bibitem{Barr}
  A.~J.~Barr,
  Phys.\ Lett.\ B {\bf 596}, 205 (2004)
  [arXiv:hep-ph/0405052],
  M.~Battaglia, A.~K.~Datta, A.~De Roeck, K.~Kong and K.~T.~Matchev,
  arXiv:hep-ph/0507284,
  A.~Datta, K.~Kong and K.~T.~Matchev,
  arXiv:hep-ph/0509246,

\bibitem{SmillieWebber}
    J.~M.~Smillie and B.~R.~Webber,
  arXiv:hep-ph/0507170,

\bibitem{IsItSUSY}
  A.~Datta, G.~L.~Kane and M.~Toharia,
  arXiv:hep-ph/0510204.

 \bibitem{NewBarr}
  A.~J.~Barr,
  arXiv:hep-ph/0511115.


\bibitem{Madgraph}
  F.~Maltoni and T.~Stelzer,
  JHEP {\bf 0302}, 027 (2003)
  [arXiv:hep-ph/0208156],
    T.~Stelzer and W.~F.~Long,
  Comput.\ Phys.\ Commun.\  {\bf 81}, 357 (1994)
  [arXiv:hep-ph/9401258],
    H.~Murayama, I.~Watanabe and K.~Hagiwara,
KEK-91-11


\bibitem{PDF}
  J.~Pumplin, D.~R.~Stump, J.~Huston, H.~L.~Lai, P.~Nadolsky and W.~K.~Tung,
  JHEP {\bf 0207}, 012 (2002)
  [arXiv:hep-ph/0201195].


\bibitem{Beneke:2000hk}
  M.~Beneke {\it et al.},
  arXiv:hep-ph/0003033.

\bibitem{Meff}
  I.~Hinchliffe, F.~E.~Paige, M.~D.~Shapiro, J.~Soderqvist and W.~Yao,
  Phys.\ Rev.\ D {\bf 55}, 5520 (1997)
  [arXiv:hep-ph/9610544].


\bibitem{ATDR}
ATLAS Collaboration TDR Vol. 1. CERN-LHCC-99-14,
ATLAS Collaboration TDR Vol. 2. CERN-LHCC-99-15



 \bibitem{MT2}
    C.~G.~Lester and D.~J.~Summers,
  Phys.\ Lett.\ B {\bf 463}, 99 (1999)
  [arXiv:hep-ph/9906349],
    A.~Barr, C.~Lester and P.~Stephens,
  J.\ Phys.\ G {\bf 29}, 2343 (2003)
  [arXiv:hep-ph/0304226].

\bibitem{ATLAStrigger}
T. Sch\"orner-Sadenius, S. Tapprogge, et. al.
ATLAS Note ATL-DAQ-2003-004.

\bibitem{ATLASstop}
G.~Polesello, L.~Poggioli, E.~Richter-W\c{a}s, J.~S\"oderqvist,
ATLAS Note PHYS-No-111, October 1997.


\bibitem{ttz}
  U.~Baur, A.~Juste, D.~Rainwater and L.~H.~Orr,
  arXiv:hep-ph/0512262.


\bibitem{PDG}
  S.~Eidelman {\it et al.}  [Particle Data Group],
  Phys.\ Lett.\ B {\bf 592}, 1 (2004).


\bibitem{ATLAStau}
D.~Cavalli and S.~Resconi,
ATLAS Note ATL-PHYS-98-118 (1998).


\bibitem{PGS}
Pretty Good Simulator,\\
\verb\http://www.physics.ucdavis.edu/~conway/research/software/pgs/pgs.html\ \\
Tuned for the LHC as per the LHC Olympics version.

\bibitem{Pythia}
  T.~Sjostrand, L.~Lonnblad, S.~Mrenna and P.~Skands,
  arXiv:hep-ph/0308153.

\bibitem{StauLHC}
R. Arnowitt et. al.,
\verb\http://hepr8.physics.tamu.edu/hep/stau/susytalkd.pdf\


\bibitem{Alpgen}
  M.~L.~Mangano, M.~Moretti, F.~Piccinini, R.~Pittau and A.~D.~Polosa,
  JHEP {\bf 0307}, 001 (2003)
  [arXiv:hep-ph/0206293].

\bibitem{lotsadegeneracies}
  N.~Arkani-Hamed, G.~L.~Kane, J.~Thaler and L.~T.~Wang,
  arXiv:hep-ph/0512190.

\bibitem{BattagliaCLIC}
  M.~Battaglia, A.~Datta, A.~De Roeck, K.~Kong and K.~T.~Matchev,
  JHEP {\bf 0507}, 033 (2005)
  [arXiv:hep-ph/0502041].

\bibitem{Minuit}
  F.~James and M.~Roos,
  Comput.\ Phys.\ Commun.\  {\bf 10}, 343 (1975).

\bibitem{LHCupgrade}
  F.~Gianotti {\it et al.},
  Eur.\ Phys.\ J.\ C {\bf 39}, 293 (2005)
  [arXiv:hep-ph/0204087].

\end{thebibliography}
\end{document}